\documentclass[12pt]{article}

\pdfoutput=1

\usepackage{amsmath,amssymb,amsfonts,amscd,mathrsfs}
\usepackage{xcolor}
\definecolor{darkblue}{rgb}{0.1,0.1,.7}
\usepackage[colorlinks, linkcolor=darkblue, citecolor=darkblue, urlcolor=darkblue, linktocpage]{hyperref} 
\usepackage{geometry}
\usepackage{tablefootnote,colortbl}

\usepackage{braket}

\usepackage{cancel}
\usepackage{arcs}
\usepackage{booktabs}

\usepackage{arydshln}
\usepackage{booktabs,multirow}
\usepackage{amssymb}
\usepackage{fancyvrb}

\usepackage{tikz-feynman}
 \tikzfeynmanset{compat=1.0.0}
 
 \usepackage[square, comma, compress,numbers]{natbib}
\usepackage{subcaption}

\definecolor{myorange}{RGB}{199,146,32}

\definecolor{Gray1}{gray}{0.97}
\definecolor{Gray2}{gray}{0.9}
\definecolor{LightCyan}{rgb}{0.88,1,1}
\definecolor{blu}{rgb}{0,0,1}

\usepackage{ragged2e}
\usepackage{array}
\newcolumntype{L}[1]{>{\raggedright\let\newline\\\arraybackslash\hspace{0pt}}m{#1}}
\newcolumntype{C}[1]{>{\centering\let\newline\\\arraybackslash\hspace{0pt}}m{#1}}
\newcolumntype{R}[1]{>{\raggedleft\let\newline\\\arraybackslash\hspace{0pt}}m{#1}}

\usepackage{dsfont} 
\usepackage{tcolorbox}
\usepackage{booktabs}

\usepackage{titlesec}
\titleformat*{\section}{\large\bfseries}
\titleformat*{\subsection}{\normalsize\bfseries}
\titleformat*{\subsubsection}{\normalsize\it}
\titleformat*{\paragraph}{\normalsize\bfseries}
\titleformat*{\subparagraph}{\normalsize\bfseries}

\def\d{\delta}

\newcommand{\beq}{\begin{equation}} 
\newcommand{\eeq}{\end{equation}}

\def\calN {{\cal N}}

\def\geq{\geqslant}
\def\leq{\leqslant}
\newcommand{\diffop}[2]{\ifthenelse{\equal{#2}{1}}{\frac{\mrm{d}}{\mrm{d} #1}}{\frac{\mrm{d}^#2}{\mrm{d} #1^#2}}}

\newcommand{\mrm}[1]{{\mathrm #1}}

\newcommand{\im}{\text{Im}\, }
\newcommand{\re}{\text{Re}\, }

\usepackage[normalem]{ulem}

\def\d{\partial}

\newcommand{\be}{\begin{equation}}
\newcommand{\ee}{\end{equation}}
\def\bea#1\eea{\begin{align}#1\end{align}}

\usepackage{accents}
\newlength{\dhatheight}

\numberwithin{equation}{section}
\usepackage{tocloft}
\begin{document}

\vspace*{-.6in} \thispagestyle{empty}
\begin{flushright}
\end{flushright}
\vspace{1cm} {\Large
\begin{center}
\textbf{QCD Worldsheet Axion from the Bootstrap}
\end{center}}
\vspace{1cm}
\begin{center}

{\bf Adwait Gaikwad$^{a}$, Victor Gorbenko$^{b}$, Andrea L. Guerrieri$^{c,d}$} \\[1cm] 
 { 
 $^a$   School of Physics and Astronomy, Tel Aviv University, Ramat Aviv 69978, Israel  \\ 
 $^b$   Laboratory for Theoretical Fundamental Physics, EPFL, Rte de la Sorge, CH-1015, Lausanne\\
 $^c$ Dipartimento di Fisica e Astronomia, Universita degli Studi di Padova, \& Istituto Nazionale di Fisica Nucleare, Sezione di Padova, via Marzolo 8, 35131 Padova, Italy. \\
 $^d$ Perimeter Institute for Theoretical Physics, Waterloo, Ontario N2L 2Y5, Canada
 }
\vspace{1cm}

\abstract{ 

The worldsheet axion plays a crucial role in the dynamics of the Yang-Mills confining flux tubes. According to the lattice measurements, its mass is of order the string tension and its coupling is close to a certain critical value. 
Using the S-matrix Bootstrap, we construct non-perturbative $2\to 2$ branon scattering amplitudes which also feature a weakly coupled axion resonance with these properties. We study the extremal bootstrap amplitudes in detail and show that the axion plays a dominant role in their UV completion in two distinct regimes, 
in one of which it cannot be considered a parametrically light particle. We conjecture that the actual flux tube amplitudes exhibit a similar behavior.

\bigskip
\noindent

}

\vspace{3cm}
\end{center}

 \vfill

\newpage 

\setcounter{tocdepth}{2}

{
\tableofcontents
}

\section{Introduction}
The description of QCD as a theory of strings remains an infamous open problem in theoretical physics. While we have accumulated an increasing amount of analytic, numerical and experimental evidence that such description is possible, at least in the limit of the large number of colors $N$, we are still lacking theoretical tools needed to describe the worldsheet theory of the corresponding strings, needless to say their interactions. In order to develop such tools, a natural object to study is a single long confining string, or a flux tube, stretching along one of the spacial directions in a pure Yang-Mills theory without dynamical quarks. It is a stable object, even at a finite $N$, and it defines a relativistic $1+1$ dimensional theory whose low-energy excitations are ``branons'', or Goldstone bosons of the spontaneously broken Poincar\'e symmetry \cite{Dubovsky:2012sh}. Below the confinement scale $ \Lambda_{QCD}$ an effective theory of branons is perturbative and decoupled from the four-dimensional excitations. This theory is constrained by the non-linearly realized Lorentz transformations which do not leave the straight string invariant. There is of course a very intuitive geometric picture of branons as describing the bending of the string in the transverse directions. The low energy action for branons looks like:
\be
S_{b}=\int d^2\sigma\left[ -{1\over 2}\left(\d_\alpha X^i\right )^2 {-}{\ell_s^2\over 8}\left (\d_\alpha X^i\d^\alpha X^i\right)^2{+}{\ell_s^2\over 4}\d_\alpha X^i\d_\beta X^j\d^\alpha X^j\d^\beta X^j+O \left (\ell_s^4\d^6 X^6 , \ell_s^6\d^8 X^4\right) \right ] .
\label{ActionBran}
\ee
Here the string length $\ell_s\sim \Lambda_{QCD}^{-1}$ is the strong coupling scale.  At finite $N$, the UV completion is given by the full YM theory containing all bulk excitations, however, when $N$ goes to infinity, it is very plausible that the $1+1$ theory is UV complete on its own. Note that there is no allowed term at the order $ O\left (\d^6 X^4\right)$, thus extending universality of the EFT to the one-loop order. This theory is what one would like to call the worldsheet theory of QCD string, from which the stringy description of QCD could be built. There are several methods that are being explored for studying this theory beyond EFT: lattice QCD, solvable toy-models, S-matrix bootstrap, and holography.

In this paper, we report on some developments in the S-matrix bootstrap approach. The flux-tube S-matrix bootstrap was initiated in \cite{EliasMiro:2019kyf}, and further developed in \cite{EliasMiro:2021nul}. One concrete motivation for our work is to understand certain numerical observation that we will call the ``triple coincidence''. This observation is related to the spectrum of massive excitations of QCD flux tubes beyond those fixed by symmetries, more precisely to the properties of the so-called worldsheet axion. The worldsheet axion\footnote{Not to be confused with the four-dimensional QCD axion.} is a particle which lives on the worldsheet and that couples to branons through a topological density that counts the self-intersection number of a string:
\be
S_{a}=\int d^2\sigma\left[-{1\over 2}\left(\d_\alpha a \right)^2-{1\over 2} m_a a^2-\ell_s^2 Q_a a\varepsilon^{ij}\varepsilon^{\alpha\beta}\d_\alpha \d_\gamma X^i\d_\beta \d^\gamma X^j+\ldots \right].
\label{ASA}
\ee
As evident from the coupling, the axion is odd under both worldsheet and target space parities, hence the name.
Lattice data unambiguously confirms the existence of the worldsheet axion, and, so far, does not hint at the presence of any other massive excitations of the QCD string. There is also an integrable toy-model, called Axionic String~\cite{Dubovsky:2015zey,Donahue:2018bch} which contains a massless worldsheet axion, and, in addition to it, we have a region in the parameter space in which the bootstrap predicts the axion as well. In this regard, the holographic models stay out from the other three approaches, as they generically predict a parity even scalar excitation, the dilaton, which parametrizes the excitation of the string in the radial AdS direction \cite{Polyakov:1998ju}. There is a conjecture that the axion may have a simple origin from perturbative QCD point of view: it matches the quantum numbers of the gluon field strength with transverse indices inserted in the Wilson line creating the flux tube on the lattice \cite{Dubovsky:2018vde}.  If the conjecture were true, the axion would be a crucial step in connecting the stringy description to asymptotic freedom of QCD. 
 
Let us now review the relation between lattice, Axionic String and bootstrap in some more details and explain the above-mentioned coincidence. 

\subsection{Worldsheet axion and the triple coincidence}
We start by briefly summarizing the lattice data on QCD flux tubes in 3+1 dimensions \cite{Athenodorou:2010cs,Athenodorou:2021vkw,Dubovsky:2013gi,Dubovsky:2014fma}. Lattice measures the energy spectrum of flux tubes of finite length winding around a compact direction.\footnote{One can also simulate open strings ending on heavy color sources. Corresponding data \cite{Juge:2002br} shows the presence of the axion as well. } Of main interest for us is the spectrum of two-particle states corresponding to two branons moving in the opposite directions. This spectrum can be related to the scattering matrix of branons and matched to the EFT lagrangian. The axion then appears as a narrow resonance in the parity odd channel of the scattering, hence lattice allows to measure with axion mass and coupling, leading to the values as in Table  \ref{tab:tab1}. The large-N value quoted here is somewhat below the one in \cite{Athenodorou:2017cmw} since in this reference the winding corrections, and other corrections caused by interactions, were not included. 
\begin{table}[t!]
\centering
 \begin{tabular}{ | c  | c | c | c |}
   \hline			
     & $SU(3)$ & $SU(5)$  & $SU(\infty)$\\
    \hline
  $2^{++}$ &  & &   \\
  $m_a^L\ell_s$ & $1.85^{+0.02}_{-0.03}$ & $1.64^{+0.04}_{-0.04}$ & 1.5\\[6pt]
   $Q_a^L$ & $0.380^{+0.006}_{-0.006}$ & $0.389^{+0.008}_{-0.008}$& -\\ [6pt]
   \hline  
   $2^{+-}$ &  &  & \\
  $m_a^L\ell_s$ & $1.85^{+0.02}_{-0.02}$ & $1.64^{+0.04}_{-0.04}$&1.5\\[6pt]
   $Q_a^L$ & $0.358^{+0.004}_{-0.005}$ & $0.358^{+0.009}_{-0.009}$& -\\ [6pt]
  \hline  
 \end{tabular}
    \caption{Axion mass and axion charge extracted from the lattice data in \cite{Athenodorou:2010cs,Athenodorou:2017cmw}. Table is taken from \cite{Dubovsky:2015zey}. The largest systematic uncertainty affecting the charge determination is coming from the lattice-induced split between two components of the symmetric-traceless flux tube state as seen from the two rows of the table. The large-N value reported here is approximate and is a simple $c_1+c_2/N^2$ fit to $SU(3)$ and $SU(5)$ points.}
      \label{tab:tab1}
\end{table}

Let us now introduce the Axionic String \cite{Dubovsky:2015zey}. The absence of one-loop counter-terms in the EFT Lagrangian leads to the universal particle production in the Goldstone sector~\cite{Dubovsky:2012sh} breaking integrability of the worldsheet theory. In the language of fundamental strings, breaking of integrability is related to the conformal anomaly which is present away from the critical dimension.\footnote{There is an alternative and equivalent approach to the effective string theory where the relation between integrability and conformal anomaly is more manifest \cite{Polchinski:1991ax,Hellerman:2014cba}.} It would be nice to have an integrable theory, to which the QCD string is in some sense close, as a toy-model to study. The only possibility to do so is to introduce another massless particle on the worldsheet, however, lattice shows that no such particles are present at least in the pure YM theory. There is, instead, a relatively light particle -- the axion. Turns out that if we take the axion exactly massless, there is an integrable theory which respects the non-linear Poincare symmetry. This theory demands that the axion coupling is equal to $Q_a^c=\sqrt{\frac{21}{48 \pi}}$,  surprisingly close to the lattice results. This coincidence was noticed in \cite{Dubovsky:2015zey} and may hint to approximate integrability at high energies, above the axion mass. In this model, the axion in some sense cancels the anomaly, even though the Axionic string is only understood in the long-string sector at the moment. It would be interesting to find a corresponding description of short strings.

The logic here seems a bit flawed: axion mass is of the order of the EFT cutoff, so why would we trust any numerology at the energies above it? In other words, the range of energies $m_a\ll E\ll \ell_s^{-1}$ does not seem to exist. One important result of our paper is to show that bootstrap provides some justification for using the approximation $m_a\ell_s \ll 1$ even for the actual axion mass.

\begin{figure}[t]
\centering
  \includegraphics[scale=.322]{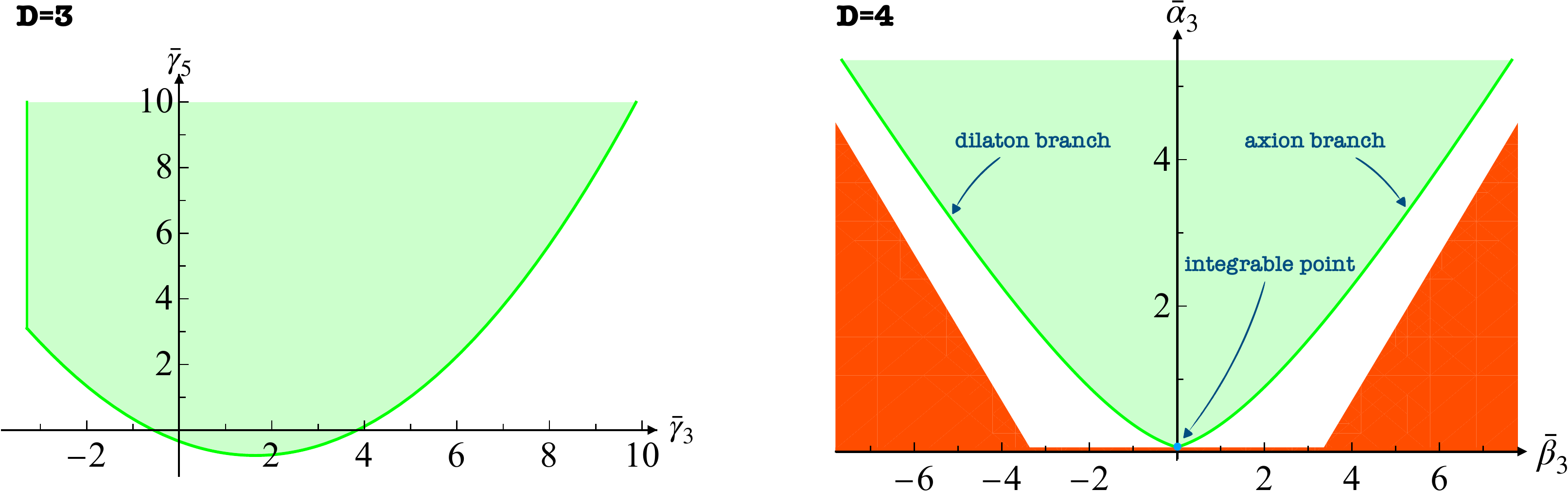}
  \caption{Left: in green, the region in the $\gamma_3-\gamma_5$ space allowed by the bootstrap constraints for $D=3$ flux tubes. Right: in green, the region in the $\beta_3-\alpha_3$ space allowed by the bootstrap constraints for $D=4$ flux tubes; in red, the region excluded analytically using the Schwarz-Pick theorem~\cite{EliasMiro:2019kyf}. With the bar we denote S-matrix coefficients rescaled as $\bar\alpha_n=2^{n+1} (4\pi)^{n-1}\alpha_n$, and the same for $\beta$ and $\gamma$. See below for the definition of these coefficients.}
\label{fig:Main_3D-4D}
\end{figure}

The numerical S-matrix bootstrap carves out the space of allowed EFTs satisfying the low energy conditions that we input, as well as non-perturbative unitarity, crossing and analyticity conditions. 
In the flux tube context this amounts, in the simplest case, to constraining the allowed values of two leading non-universal coefficients in the $2\to2$ worldsheet S-matrix. These coefficients, which appear at the order $\mathcal{O}(\d^8 X^4)$ in the action are denoted by $\alpha_3$ and $\beta_3$. The allowed region has the form as shown in figure~\ref{fig:Main_3D-4D}, right.
Using the bootstrap algorithm it is also possible to reconstruct and study the extremal amplitudes along the boundary of the allowed region.
As was first observed in \cite{EliasMiro:2019kyf} there are two pronounced boundaries, and one of them has a single sharp resonance which is an axion.\footnote{The other boundary shows the presence of a sharp dilaton instead of the axion. At $D=4$ there exists a symmetry of the crossing constraints under the exchange of the singlet and the antisymmetric channels. This transformation also sends $\beta_3\to -\beta_3$, leaving $\alpha_3$ invariant.} Mass of the axion varies along the boundary, and both $m_a=m_a^L$ and $m_a\approx0$ are included. Once we fix the mass, position on the boundary is determined and the coupling of the axion can be determined giving a value $Q_a^{b}$. It happens to be again close to $Q^c_a$ \cite{EliasMiro:2019kyf}. We thus have the triple coincidence:
\be
Q_a^L\approx Q_a^c \approx Q_a^{b}
\label{triple}
\ee

In this paper, we improve the numerical data on the flux tube S-matrix bootstrap, focusing on the determination of the properties of the axion. We significantly improve the numerical precision which is made possible by rescaling the kinematic variables with the Wilson coefficients in an appropriate way. We will observe that in the asymptotic region of large positive $\beta_3$, the axion mass becomes small and the EFT which explicitly contains the light axion gives a good analytic approximation to the results. In this regime, the coincidence $Q_a^c \approx Q_a^{b}$ can be readily explained. In both cases, the axion coupling is fixed so that it cancels the $O(s^2)$ growth of the phase shift. The regime of validity of the EFT roughly corresponds to an axion mass smaller than the large-$N$ value. We will see, however, that numerically the coincidence still holds even away from the validity of this EFT, in particular when the axion mass is close to its $SU(3)$ value. In this regime, we have a less rigorous explanation for it which is based on dispersion relations. If one assumes that the UV completion of the real flux tube theory is not very different from the boundary bootstrap S-matrix, it would also explain the first equality in \eqref{triple}. Flipped another way, one can take the coincidence as an evidence for the similarity of UV completions. Namely, our results provide additional evidence that existence of the axion is related to non-criticality of the QCD string.

While this paper is focused on strings in four-dimensional space, we also discuss the cases $D=3$ and $D>4$. In particular, an analog of the axionic string can be considered where the axion is replaced by the anti-symmetric form field: $a\varepsilon^{ij}\to A^{ij}$ \cite{Dubovsky:2015zey}. Theories with massive anti-symmetric form fields are also found by bootstrap. Unfortunately we do not have any confining gauge theories to compare with. Also, self-interactions of the form-fields imposed by non-linear Lorenz symmetry make them harder to construct \cite{Donahue:2018bch}.  In $D=3$, lattice does not predict any resonances, in some sense one can think of it being consistent with continuation of the axionic string to $D=3$ since anti-symmetric form has no components. Bootstrap results are summarized on figure \ref{fig:Main_3D-4D}, and we will explain their interpretation below.

\section{Scaling theory for light resonances}

In this section, we explore the EFTs with light resonances that provide an analytic understanding of some boostrap bounds. In addition to this, they help to improve significantly the numerical convergence of our procedure.

Numerical S-matrix bootstrap is a non-perturbative tool, and often we do not have a good analytic understanding of the theories that lie on the boundary of the allowed region. We expect, however, that at least in various asymptotic regions such understanding is possible.~\cite{Guerrieri:2020bto, Acanfora:2023axz} In particular, we will study the phenomenologically relevant part of the boundary with large and positive $\beta_3$. It will turn out that a simple EFT containing a light worldsheet axion explains the leading scaling of S-matrix coefficients in this limit. This will be the key observation that will allow us to connect the bootstrap results with the Axionic String model. Before we discuss the relevant 4D case, let us first study the simpler 3D case in this language.

\subsection{$3D$}
 
In $D=3$, the phase shift of the $2\to 2$ scattering amplitude of branons can be described at low energies by a leading universal term proportional to $s \ell_s^2$, and by a number of non-universal sub-leading corrections, characterized by dimensionless parameters $\gamma_n$
\be
2\delta(s)=\frac{s}{4}\ell_s^2+\gamma_3 s^3 \ell_s^6+\gamma_5 s^5 \ell_s^{10}+\gamma_7 s^7 \ell_s^{14}+i \gamma_8  s^8 \ell_s^{16}+\dots
\ee
The phase shift is real and the scattering elastic up to $\mathcal{O}(s^8 \ell_s^{16})$, when the inelasticity due to $2\to 4$ processes kicks in.
The leading particle production coefficient is not arbitrary, but related to the first non-universal correction to Nambu-Goto $\gamma_3$ by the non-linearly realized Poincare symmetry
\be
\gamma_8=\frac{81}{2^{15} \times 7^2 \times \pi^2}\gamma_3^2.
\ee
The first three non-universal corrections $\{ \gamma_3, \gamma_5, \gamma_7\}$ satisfy a set of analytic inequalities derived assuming that the low energy EFT amplitude is the expansion of a generic nonperturbative amplitude satisfying maximal analyticity, crossing, unitarity, and polynomial boundedness\footnote{More precisely we just need the amplitude to grow slower than exponential.}
\begin{eqnarray}
&&\tilde \gamma_3 \geq 0 \implies \gamma_3 \geq -\frac{1}{768}\simeq -\frac{3.3}{(4\pi)^2\times 2^4},\nonumber\\
&&\tilde\gamma_5 \geq 4 \tilde\gamma_3^2-\frac{\tilde\gamma_3}{64} \geq 0 \implies \gamma_5 \gtrsim -\frac{4.9}{(4\pi)^4 \times 2^6}, \nonumber\\
&&\tilde\gamma_7\geq \frac{\tilde\gamma_5^2}{\tilde\gamma_3}+\frac{\tilde\gamma_5}{64}-\frac{\tilde\gamma_3^2}{16}+\frac{\tilde\gamma_3}{4096}\geq 0 \implies \gamma_7 \gtrsim -\frac{2.1}{(4 \pi)^6\times 2^{8}},
\label{SchwarzPick}
\end{eqnarray}
where
\be
\tilde\gamma_n=\gamma_n+(-1)^{\frac{n+1}{2}} \frac{1}{n 2^{3n-1}}.
\ee
Those inequalities are saturated by integrable $S$-matrices.
For instance, the optimal value of $\gamma_3$ is saturated by an $S$-matrix with a single broad resonance with mass $m=(2+2 i)\ell_s^{-1}$ 
\be
S_\text{cusp}=-S_\text{Goldstino}=\frac{8 i-s \ell_s^2}{8 i+s \ell_s^2}.
\ee
The inequality for $\gamma_5$ is saturated by a family of $S$-matrices with one complex resonance with a mass depending on $\gamma_3$ 
\be
S_\text{edge}=\frac{\hat s^2 \tilde \gamma_3-192 -24 i \hat s}{\hat s^2 \tilde \gamma_3-192 +24 i \hat s}, \quad \hat s=s \ell_s^2.
\label{S_edge}
\ee

We can determine the large $\gamma_3$ behavior of this resonance from the exact solution to find
\be
m \ell_s=\frac{1}{\sqrt{2}\gamma_3^{1/4}}+\frac{i}{64\sqrt{2}\gamma_3^{3/4}}+\mathcal{O}(\gamma_3^{5/4})
\label{mass_3d}
\ee
At large $\gamma_3$, this resonance becomes light and weakly coupled. 
In particular, the cutoff of the low energy EFT of branons is then only set by the mass of this resonance. 
As we will explain momentarily, we could imagine describing the asymptotic bootstrap bounds for large $\gamma_3$ by using a simple tree level amplitude computed in the EFT with the light particle. 

Consider the simple Lagrangian describing the interaction of the branons $X$ with a massive scalar field $\phi$
\be
\mathcal{L_\phi}=-\frac{1}{2}(\partial X)^2-\frac{1}{2}(\partial \phi)^2-\frac{1}{2}m^2 \phi^2-\frac{1}{2}g (\partial X)^2\phi.
\ee
Of course this Lagrangian does not respect the non-linear Lorentz invariance, however, bootstrap is also blind to it, so for our current purposes we can use it.
The tree-level scattering amplitude $XX \to XX$ is
\be
M^\text{tree}(s)=-\frac{g^2 m^4}{4}\left(\frac{1}{s-m^2}-\frac{1}{s+m^2}\right)-\frac{g^2 m^2}{2}.
\ee
At energies $s\ll m^2$, we simply have
\be
M^\text{tree}(s)=\frac{g^2 m^2}{2}\left(\frac{s^2}{m^4}+\frac{s^4}{m^8}+\frac{s^6}{m^{12}}+\frac{s^8}{m^{16}}+\mathcal{O}(s^{10}) \right).
\label{UV model}
\ee

In order to make predictions we need to match the above expansion with the low energy expansion of the flux-tube amplitude in the large $\gamma_3$ limit.
The generic low energy expansion of the flux-tube amplitude takes the form
\be
M^\text{EFT}(s>0)=-2i s(S-1)=\frac{s^2\ell_s^2}{2}+i \frac{s^3 \ell_s^4}{16}+(2\gamma_3-\tfrac{1}{192})s^4 \ell_s^6+\dots
\label{EFT3D}
\ee
This expansion is valid when $s \ell_s^2 \ll1$. When $\gamma_3$ is large, the mass of the resonance becomes parametrically smaller than $\ell_s^{-1}$, namely, $m\ell_s \sim \gamma_3^{-1/4}$.

Matching the first two terms in \eqref{UV model} and the first and third term in \eqref{EFT3D} yields in the large $\gamma_3$ limit
\be
m\ell_s=\frac{1}{\sqrt{2} \gamma_3^{1/4}}, \quad g=\frac{1}{\sqrt{2} \gamma_3^{1/4}}.
\label{predicted_scaling}
\ee
Note that the terms proportional to the odd powers of $s$, naturally associated to loops corrections, are suppressed by powers of $\gamma_3$, while for the even powers we get 
\be
\gamma_5=4 \gamma_3^2+\dots, \quad \gamma_7=16 \gamma_3^3+\dots
\ee
that matches the asymptotic behavior in eq.~\eqref{SchwarzPick}.

So far, we exploited our model to match tree-level quantities, but we can go a bit further with the use of perturbative unitarity to compute the imaginary part of the $\mathcal{O}(g^4)$ amplitude
\be
2\im M=\frac{1}{2s}|M^2|\equiv \frac{1}{2s} (M^\text{tree})^2
\ee 
that matches perfectly the order $\mathcal{O}(s^3)$ term in eq.~\eqref{EFT3D}.
The tree-level matching correctly predicts the leading large $\gamma_3$ behavior of the mass of the resonance.
To access the sub-leading term we need to compute the one-loop correction to the self-energy, alternatively we can use the Inverse Amplitude Method~\cite{Truong:1988zp,Dobado:1996ps} and take the tree level phase shift to write a unitarized ansatz of our tree level model
\be
S^\text{U}(s)=\frac{1+i\delta^\text{tree}(s)}{1-i \delta^\text{tree}(s)}=1+2 i \delta^\text{tree}(s)-2(\delta^\text{tree}(s))^2+\dots
\label{unitarized_S}
\ee
where by $\delta^\text{tree}(s)$ we mean the leading phase shift in $g^2$
\be
\delta^\text{tree}(s)=\frac{M^\text{tree}(s)}{4 s}.
\ee
We use \eqref{unitarized_S} as a resummed expression, which is equivalent to resumming 1PI propagator for $\phi$. Resonance then corresponds to a zero of this object.
\be
1+i \delta^\text{tree}(s)=0 \implies s=m^2(1+i\tfrac{g^2}{16}+\mathcal{O}(g^4)),
\ee
that matches eq.~\eqref{mass_3d} after replacing the mass and the coupling with eq.~\eqref{predicted_scaling}.

As a last comment, we check that using the rescaling $s \ell_s^2=\frac{\bar s}{\sqrt{\gamma_3}}$ it is possible to obtain a systematic large $\gamma_3$ expansion of the exact $S_\text{edge}$ solution
\be
S_\text{edge}(\bar s)=1+\frac{i}{4\sqrt{\gamma_3}}\frac{\bar s}{1-4\bar s^2}-\frac{1}{32 \gamma_3}\frac{\bar s^2}{(1-4\bar s^2)^2}+\dots
\ee
that matches our tree level model.

\subsection{$4D$}
\label{sec:EFT4D}
Interpretation of bootstrap results is more complicated in 4D since we do not have a simple analytic expression for the boundary S-matrix. Let us first summarize the numerical bootstrap results. 
The phase shifts at low energies behave respectively for the singlet, antisymmetric, and symmetric channel as
\begin{eqnarray}
&&2\delta_\text{sing}(\hat s)=\frac{\hat s}{4}-\alpha_2 \hat s^2+(\alpha_3-2\beta_3)\hat s^3-\alpha_4 \hat s^4+(\alpha_5-2\beta_5-\tfrac{4\alpha_2 \beta_3}{\pi}\log(\hat s))\hat s^5+\dots\nonumber\\
\label{4D_ps}
&&2\delta_\text{anti}(\hat s)=\frac{\hat s}{4}-\alpha_2 \hat s^2+(\alpha_3+2\beta_3)\hat s^3-\alpha_4 \hat s^4+(\alpha_5+2\beta_5+\tfrac{4\alpha_2 \beta_3}{\pi}\log(\hat s))\hat s^5+\dots\\
&&2\delta_\text{sym}(\hat s)=\frac{\hat s}{4}+\alpha_2 \hat s^2+\alpha_3\hat s^3+\alpha_4 \hat s^4+\alpha_5\hat s^5+\dots\nonumber
\end{eqnarray}
with $\alpha_2=-\frac{22}{384\pi}$ fixed by non-linearly realized Poincar\'e symmetry. This term is sometimes referred to as the Polchinski-Strominger (PS) term. Note that it does not correspond to any term in the action for branons, it is a result of the one-loop calculation with the leading interaction vertices in \eqref{ActionBran} \cite{Dubovsky:2012sh}.
Analytically, we know that 
\be
\alpha_3 \geq -\frac{1}{768}+\frac{121}{9216 \pi^2}, \quad \alpha_3 \geq -\frac{1}{768}+|\beta_3|.
\ee
In $D=4$, the crossing constraints are symmetric with the exchange of singlet and antisymmetric channel.
Numerically, we observe that the bootstrap bound shows the presence of a pronounced resonance in the anti-symmetric channel for $\beta_3>0$ and in the singlet channel for $\beta_3<0$. We refer to these resonances as worldsheet axion and dilaton respectively, as we already discussed in the introduction. 

Let us focus on the axion case. Numerically, we observe that when $\beta_3>0$ is large, the axion mass scales like $m_a\ell_s \sim \beta_3^{-1/2}$ (see section \ref{} for details). Thus, similarly to the 3D case the resonance becomes light, however, the scaling is different compared to \eqref{mass_3d}. This suggests to try an effective theory with a different number of derivatives in the coupling:
\be
\mathcal{L}_a=-\frac{1}{2}(\partial X^i)^2-{1\over 2}\left(\d a \right)^2-\frac{1}{2}m_a^2 a^2-g_a a \varepsilon_{ij}\varepsilon^{\alpha\beta} \d_\alpha \d_\gamma X^i \d_\beta \d^\gamma X^j+\dots
\label{La}
\ee
We use a different coupling than in \eqref{ASA} to emphasize that so far the theories are not related.

A simple tree-level computation produces
\begin{eqnarray} 
&&M_\text{sing}=-\frac{g_a^2}{4}\frac{s^4}{(s+m_a^2)},\nonumber\\
&&M_\text{anti}=-\frac{g_a^2}{4}\frac{s^4(s+3m_a^2)}{(s-m_a^2)(s+m_a^2)},\\
&&M_\text{sym}=\frac{g_a^2}{4}\frac{s^4}{(s+m_a^2)}. \nonumber
\label{Maxion}
\end{eqnarray}

We see that the expansion of these amplitudes at low energies starts from the $\alpha_3$ and $\beta_3$ terms, while the tree-level and one-loop NG terms are not produced by it. We thus match the leading order expansion of the axion amplitudes for $s\ll m^2$ with the $\mathcal{O}(\hat s^3)$ of eq.~\eqref{4D_ps} 
\be
\frac{g_a^2}{8 m_a^2}\{-1,3,1 \}=\ell_s^6\{(\alpha_3-2\beta_3),(\alpha_3+2\beta_3),\alpha_3\} \implies \{\alpha_3=\beta_3, g_a=2\sqrt{2\beta_3} m_a \ell_s^3\}.
\ee
We then match the next order term, which is also real in this case, to the
the Wilson coefficient $\alpha_4$. This leads to
\be
\alpha_4= -a_4 \beta_3^2, \quad \text{with}, \qquad m_a \ell_s=\frac{1}{\sqrt{a_4 \beta_3}}, \quad g_a=\frac{2\sqrt{2} \ell_s^2}{\sqrt{ a_4}},
\label{relations}
\ee
 where $a_4$ is positive and it does not scale with $\beta_3$.
To summarize, we related the coefficients $\alpha_3$, $\beta_3$ and  $\alpha_4$ to the mass and the coupling of the axion. Three parameters are matched to two because of the constraint $\alpha_3=\beta_3$, satisfied asymptotically by the bootstrap bound, and consistent with the weakly coupled axion EFT. Our axion EFT coupling $g_a$ is dimensionful, so it is a weakly coupled particle at the scale of its own mass as long as $m_a \ell_s \ll1$ is satisfied. 

Using the tree level axion amplitude, we also predict the following relations between higher order Wilson coefficients: 
\be
\alpha_5=\beta_5=a_4^2 \beta_3^3, \quad \alpha_6=-a_4^3 \beta_3^4, \quad \ldots
\ee

So far the axion EFT appears to be unrelated to the string tension which controls the leading NG interactions.\footnote{ $\ell_s$ appears in  \eqref{relations} only because we chose to express the higher order S-matrix coefficients in the units of $\ell_s$. This was done in anticipation of the correct scaling, but we could have used any other scale so far.} How does this relation come about? The logic here is different from what we had in 3D: instead of matching in the IR, the matching happens in the UV with respect to the axion mass. Note that at $s\gg m_a^2$, the axion amplitudes grow as $s^3$, and so does the PS, or $\alpha_2$, term. This growth of the amplitude, corresponding to the $s^2$ growth of the phase shift, is very unphysical (for example the S-matrix $e^{i c s^2}$ explodes on the physical sheet for any $c$), thus bootstrap cancels this growth by matching the axion amplitudes with the PS term. This condition reads:
\be
\frac{g_a^2}{4}=- 2 \ell_s^4 \alpha_2 = \ell_s^4 \frac{22}{192\pi} \,.
\label{UVmatch}
\ee
This implies, in turn, that the dimensionless S-matrix coefficient $a_4$ is set to be 
\be
a_4=-\frac{1}{\alpha_2}.
\ee
After this identification all S-matrix coefficients are fixed in terms of the universal EFT terms $\ell_s$, $\alpha_2$ as well as the leading non-universal coefficient $\beta_3$, to the leading order in the large-$\beta_3$ expansion.

Let us now compare the resulting axion coupling with the Axionic String prediction. Of course, the condition of UV-cancelation is the same as in the the Axionic String. The dimensionless coupling we get is, however, slightly different: 
\be
Q_a^{EFT}\approx\ell_s^{-2}g_a=\sqrt{\frac{22}{48 \pi}}
\label{Qbootstr}
\ee
To understand the difference from $Q_a^{c}=\sqrt{\frac{21}{48 \pi}}$, let us remember that the $s^3$ term in the amplitude comes form the loop of branons contributing a factor $\sim D-26=22$ \cite{Dubovsky:2012sh}. Once the axion is added to the theory, the non-linearly realized Poincar\'e symmetry demands it to also runs in the loop transforming 22 into 21. This comes form the universal $(\d a)^2(\d X^i)^2$ term in the effective action.  Bootstrap, in its current realization, is blind to the symmetry and, likely does not know about the axion running in the loop. We will come back to this point in the discussion of the numerical results.

A comment is in order. Even though we did not introduce the non-linear Poincar\'e symmetry into bootstrap, the effective theory that we picked, namely \eqref{La}, is fully consistent with it since it is equivalent to \eqref{ASA}. While the most dangerous, order $s^2$, term in the phase shift is cancelled by the light axion, the linear in $s$ growth is not affected by it at the leading order in large $\beta_3$ expansion. While the phase shift linear in $s$ is consistent with the known S-matrix properties, we expect that the optimal S-matrix does not have it, at least not in all channels. The physical flux tube phase-shift is expected to have some linear piece in the UV, however, it should be smaller than that in the IR \cite{Dubovsky:2018vde}. As we will see, bootstrap cancels this linear growth by heavy resonances with couplings not respecting non-linear Poincare, similarly to the 3D case. We thus expect that the UV part of the amplitude, parametrically above the axion mass, can differ in the bootstrap case and for the real flux tube theory, in particular it will have some particle production.

As we did in $D=3$, we can repeat the unitarization procedure and derive the asymptotic dependence of the axion width on $\beta_3, a_4$.
Using \eqref{Maxion}, \eqref{relations} and \eqref{UVmatch} we get 
\be
2\delta_\text{anti}(\hat s)=\frac{\hat s}{4 }-\alpha_2 \hat s^2+\alpha_2\hat s^3 \frac{\hat s -\frac{3 \alpha_2}{ \beta_3}}{\left ( \hat s+\frac{ \alpha_2}{ \beta_3}\right)\left ( \hat s-\frac{\alpha_2}{ \beta_3}\right)}.
\label{eq:asymptotic_anti}
\ee

If we perturbatively solve the equation near the axion mass the NG terms are suppressed, and for large $\beta_3$ we get
\be
1+i\delta_\text{anti}(\hat s_a)=0\implies \hat s_a=\frac{|\alpha_2|}{\beta_3}+\frac{i\alpha_2^4}{ \beta_3^3}+\dots
\ee
we obtain the imaginary part of the axion mass
\be
m_a \ell_s=\sqrt{\frac{|\alpha_2|}{\beta_3}}+i \frac{|\alpha_2|^{7/2}}{2 \beta_3^{5/2}}+\dots
\ee

Therefore we have a prediction for the ratio 
\be
Q_a=\frac{\sqrt{8\Gamma_a}}{m_a^{5/2}\ell_s^2}=2\sqrt{2 |\alpha_2|}+\dots
\label{Qa}
\ee

So far our discussion in this section was limited to the case of large $\beta_3$, or equivalently small $m_a \ell_s$. Just form the EFT perspective we cannot conclude whether the cancellation discussed above should occur for the physical values of the axion mass, however, bootstrap allows us to do precisely that. We can go to the point on the boundary corresponding to the physical mass and check numerically how well \eqref{UVmatch} or \eqref{Qbootstr}. works. The detailed analysis is presented below, but the short summary is that, according to bootstrap, $m_a \ell_s\ll1$ is already a decent approximation for large-N, and both equalities are satisfied with a reasonable precision. For the higher values of the mass, including the $SU(3)$ case, \eqref{UVmatch} fails, however \eqref{Qbootstr} continues to hold. It suggests the cancellation of the UV growth by the axion resonance holds more non-perturbatively, as we discuss in section \ref{sec:dominance}.  It thus likely that QCD strings also utilize the axion to cancel the PS growth, which explains the second part of the triple coincidence: approximate equality of lattice and critical axion couplings.

\section{Improved data for Flux Tube Bootstrap}
\label{section_improved_numerics}

Let us now discuss the numerical procedure that we implemented.
In any dimension $D>3$, the flux tube two-to-two scattering amplitude depends on a number of low energy constants $\{\alpha_3,\beta_3,\alpha_4,\dots \}$ that can be unambiguously related to the Wilson coefficients appearing in the EFT Lagrangian.
Assuming that the nonperturbative scattering amplitude of branons satisfies the $S$-matrix constraints of analyticity, crossing, and unitarity, we can use the Bootstrap to determine the allowed space of the EFT Wilson coefficients. 

Here we follow the constructive \emph{primal} approach introduced in \cite{Paulos:2016but}, and adapted to the flux tube branons case in \cite{EliasMiro:2019kyf}.
In the primal Bootstrap, we construct amplitudes compatible with the $S$-matrix constraints. 
The first step is to make an analytic and crossing symmetric ansatz for the $S$-matrix.
We recall that, in generic $D$, the branons are labelled by an index $i=1,\dots, D-2$ corresponding to one of the transverse directions to the flux tube.
In general
\be
\mathbb{S}_{ij}^{kl}(s)=\delta_{ij}\delta^{kl}\sigma_1(s)+\delta_{i}^k\delta_{j}^l \sigma_2(s)+\delta_{i}^l\delta_j^k \sigma_3(s)
\ee
where the three scalar functions $\sigma_a(s)$ can be restricted to take value in the upper half plane (UHP) where they satisfy the following crossing and real analyticity properties
\be
\sigma^*_2(-s^*)=\sigma_2(s), \quad \sigma^*_1(-s^*)=\sigma_3(s).
\ee

We can write an elegant and compact ansatz for these functions using the map from the UHP to the unit disk
\be
\chi_{s_0}(s)=\frac{i s_0-s}{i s_0+s}
\ee
and Taylor expand in $\chi$
\bea
\sigma_1(s)&=\sum_{n=0}^{N_\text{max}} (a_n+i b_n)\chi_{s_0}(s)^n\nonumber\\
\sigma_2(s)&=\sum_{n=0}^{N_\text{max}} c_n \chi_{s_0}(s)^n\nonumber\\
\sigma_3(s)&=\sum_{n=0}^{N_\text{max}} (a_n-i b_n)\chi_{s_0}(s)^n.
\label{ansatz_4D}
\eea
We truncate the Taylor expansion at the order $N_\text{max}$ since we want to evaluate this ansatz numerically.
For fixed $N_\text{max}$, we explore a subset of all possible amplitudes that becomes complete when $N_\text{max}\to\infty$.
The set of real coefficients $\{a_n,b_n,c_n\}$ satisfy an infinite number of quadratic inequalities derived by imposing unitarity $|S_\text{irrep}(s)|^2 \leq 1$ for any $s\geq 0$ on the irreducible representations of the $O(D-2)$ flavour group
\bea
S_\text{sing}(s)&=(D-2)\sigma_1(s)+\sigma_2(s)+\sigma_3(s)\nonumber\\
S_\text{anti}(s)&=\sigma_2(s)-\sigma_3(s)\nonumber\\
S_\text{sym}(s)&=\sigma_2(s)+\sigma_3(s).
\eea

To zoom in on the space of non-linear-Lorentz invariant amplitudes, we have to impose a number of linear low energy constraints, 
which arise from matching our Ansatz~\eqref{ansatz_4D} to 
the low energy expansion in eq. \eqref{4D_ps}. 
For instance, at leading order in the small energy expansion $\sigma_2(0)=1$, and $\sigma_1(0)=\sigma_3(0)=0$, thus we get
\be
\sum_{n=0}^{N_\text{max}} c_n=1, \quad \sum_{n=0}^{N_\text{max}} a_n=\sum_{n=0}^{N_\text{max}} b_n=0,
\ee
and similarly for the higher orders in $s$. Indeed, we implement this matching up to $\mathcal{O}(s^3)$ where the first non universal correction to Nambu-Goto appear in the low energy expansion.\footnote{Strictly speaking, to impose the soft theorems from nonlinearly realized Poincar\'e and elastic unitarity would be sufficient to match up to $\mathcal{O}(s^2)$.} 
 
 \begin{figure}[t]
\centering
  \includegraphics[scale=.267]{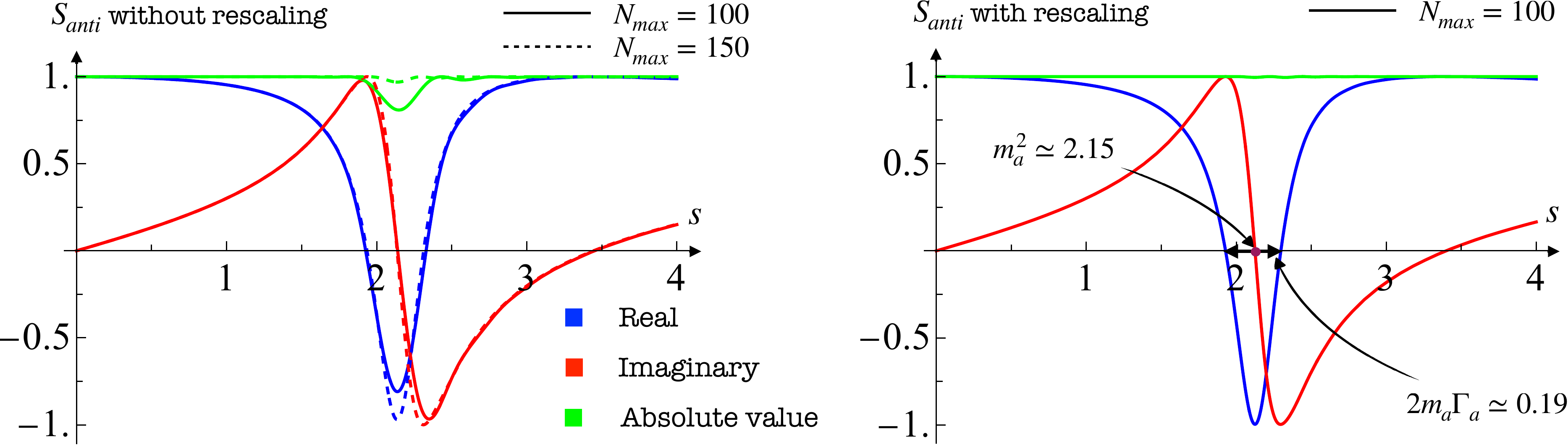}
  \caption{S-matrix in the antisymmetric channel as a function of $s$ obtained by minimizing $\alpha_3$ at fixed $\bar \beta_3=22.74$. Left: result of the optimization using an Ansatz without rescaling. Right: result using an Ansatz with the rescaling $s_0=|\alpha_2|/\beta_3$. All dimensionful quantities appearing in the figures are expressed in units of $\ell_s=1$.}
\label{fig:Numerical_Improvements}
\end{figure}
 
The Ansatz in \eqref{ansatz_4D} is the general parametrization of a holomorphic function inside the unit disk. When $N_{max}\to\infty$, it gives the Taylor series representation of any analytic function. On the other hand, not all functions can be well approximated using a truncated series representation with $N_{max}$ terms. 
A class of functions that are hard to approximate is given by weakly coupled resonances, namely zeros of the $S$-matrix, that are close to the real axis.
A weakly coupled resonance is parametrized by its mass and width obeying the condition $\Gamma\ll m$. These quantities are related to the position of the zero in the upper half $s$-plane through the formula $s_R=(m+i\tfrac{\Gamma}{2})^2\simeq m^2+i m\Gamma$. Around the position of the resonance, the real and imaginary parts behave universally forming the typical structure shown in figure~\ref{fig:Numerical_Improvements} (right panel). In figure~\ref{fig:Numerical_Improvements}, we take as a benchmark point $\bar \beta_3=(4\pi)^2 2^4 \beta_3=22.74$ that correspond roughly to the value associated to the $SU(\infty)$ flux-tube theory. At this point, the axion has mass $m_a=1.467$ and width $\Gamma_a=6.512 \times10^{-2}$. From the figure, it is clear that the resonant structure is contained in a window of size $\Delta s=2 m\Gamma$. When we map the upper half-plane to the unit disk, or equivalently the positive real axis to the upper boundary of the disk $\chi_{s_0}=e^{i\phi}$, this window will be deformed according to the simple formula
\be
\Delta \phi=\frac{2s_0}{s^2+s_0^2}\Delta s=\frac{2s_0}{m^4+s_0^2}2 m\Gamma.
\ee
Since we Taylor expand in the unit disk, we can define a critical value $N_\text{critical}=\pi/\Delta\phi$ corresponding to the minimum value of $N_\text{max}$ needed to resolve the resonance, that we empirically associate to the appearance of the zero in the complex plane.
Moreover, to have a precise reconstruction of the resonance we want to make sure that $\Delta\phi \gg \pi/N_\text{max}$ or $N_\text{max}\gg N_\text{critical}$. 
The maximum of the jacobian factor is attained when $s_0=m^2$. In that case $\Delta\phi\equiv \tfrac{2\Gamma}{m}$, which depends only on the intrinsic properties of the resonance and not on the conformal map chosen. Thus, it is clear that with a careful choice of $s_0$ we can minimize the number of terms in the ansatz needed to resolve the resonance, however there is an intrinsic lower bound on $N_\text{max}$ which depends only on the ratio $\tfrac{\Gamma}{m}$.

In general, resonances are dynamically generated by solving the bootstrap optimization problem, and it is hard to know a priori where they will appear.  However, according to the EFT predictions discussed in section \ref{sec:EFT4D} we can tune $s_0$ as a function of $\beta_3$ according to 
\be
s_0=\frac{|\alpha_2|}{\beta_3}\simeq \frac{46}{ \bar\beta_3}, \implies \Delta\phi=2 \frac{|\alpha_2|^{5/2}}{\beta_3^{3/2}}\simeq \frac{11.4}{\bar\beta_3^{3/2}}
\ee
in order to use each time the best possible ansatz to describe the dynamically generated axion in the theory. For instance, in the example in figure \ref{fig:Numerical_Improvements} left panel, we plot the $S$ matrix in the antisymmetric channel for two values of $N_\text{max}=100$ (in solid) and $N_\text{max}=150$ (in dashed). In this example we have chosen $s_0=8$, a value used for instance in \cite{EliasMiro:2019kyf}. Using our estimate we can predict a critical value $N_\text{critical}\simeq 60$. Indeed, for $N_\text{max}=100>N_\text{critical}$ the resonant structure is already well reproduced, but not perfectly converged. If we push a bit higher to $N_\text{max}=150$, then the absolute value (the green dashed line) saturates unitarity almost perfectly close to the resonance.

With the rescaling, we can set $s_0\simeq 2.05$, and predict a critical $N_\text{critical}\simeq 30$. In fig.~\ref{fig:Numerical_Improvements}, right panel, we check that using $N_\text{max}=100$ the axion resonance is already perfectly reproduced. Rescaling becomes even more relevant for larger values of $\beta_3$, and in what follows we always use numerics with rescaling.

\begin{figure}[t]
\centering
  \includegraphics[scale=.265]{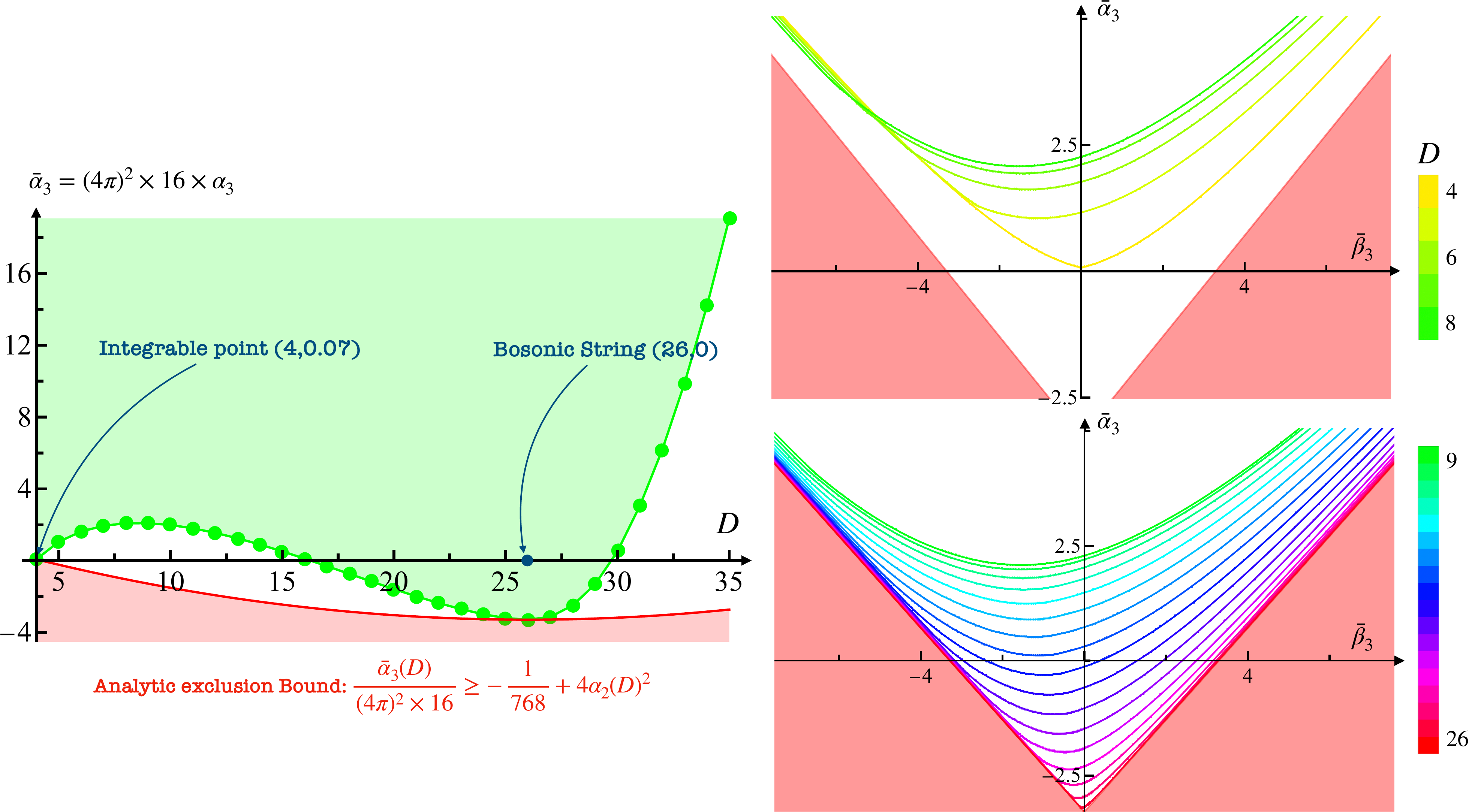}
  \caption{Left: minimum of $\bar\alpha_3$ as a function of the dimension $D$ (greed dots). We denote the allowed region of $\bar\alpha_3$ in green. In red, the region excluded by the analytic Schwarz-Pick inequality \cite{EliasMiro:2019kyf} applied to the symmetric channel S-matrix. Right: boundary of the allowed region in the $\bar\beta_3 -\bar\alpha_3$ plane for different dimensions $D$. In red, the analytically excluded region using the Schwarz-Pick inequality applied to the two crossing symmetric components $\sigma_2$ and $\sigma_1+\sigma_2+\sigma_3$ which gives $\bar\alpha_3 > -\pi^2/3 +|\bar\beta_3|$.}
\label{fig:All_D_boundary}
\end{figure}

\section{Results}
\subsection{Results in general $D$}
Using the methodology described in the previous section, we can explore the space of Flux Tube amplitudes. We will consider a simple optimization problem where we fix the value of $\beta_3$ and we minimize $\alpha_3$. The result in several dimensions larger or equal to 4 are given in Figure \ref{fig:All_D_boundary}. On the left figure we show the absolute analytic lower bound on $\alpha_3$. For $D=4, 26$ it is achieved for $\beta_3=0$, while it is not so for other dimensions. $D=26$ is special because it corresponds to the critical string. In this case, $\alpha_2$ vanishes and the kink corresponds to the integrable theory known as the $T\bar T$ deformation of 24 free bosons. $D=4$ is special because of the existence of another integrable theory identified in \cite{Cooper:2014noa}.

 On the right figure, we show the boundaries of allowed regions for $D$ up to 26. We see that the boundaries quickly approach straight lines in agreement with the scaling predicted by EFTs with light particles. We briefly describe the corresponding EFTs possessing the dilaton or the anti-symmetric form particle in Appendix \ref{App:anit-sym} and \ref{dilaton_appendix}. We expect these theories to have the same properties as the light axion EFT studied in section \ref{sec:EFT4D}.\footnote{{We have explicitly checked the EFT predictions for $\bar\beta_3\gg 1$ and $D=5, 15$. }} We did not explore in details the boundary for $D>26$, however, our preliminary analysis confirms expectations of \cite{Dubovsky:2015zey}. For $D>26$ $\alpha_2$ is negative, and one needs a particle in the symmetric representation of $SO(D-2)$ in order to cancel the amplitude growth in the UV. Such a resonance indeed does appear.

\subsection{Detailed results in $D=4$}
\label{sec:detailed_results}
We now switch to describing our results in $D=4$. In this, and in the next section, we set $\ell_s=1$, unless we explicitly write it.
In Appendix~\ref{appendix_table}, we report all the data we will study in this section.
Our main focus is on the region of large and positive $\bar\beta_3$, since it corresponds to QCD flux tubes. The boundary quickly becomes a straight line with the unit slope, in accord with our EFT predictions.

We start with the first asymptotic prediction $\bar\alpha_3=\bar\beta_3$. 
In general, we can parametrize the corrections to the tree level EFT result generated by integrating out the UV physics with an expansion in inverse powers of the form\footnote{As already done before, in this section we will rescale any Wilson coefficient appearing at order $\mathcal{O}(s^n)$ in the small energy expansion of the phase shift by a factor $2^{n+1} (4\pi)^{n-1}$ to take into account naive dimensional analysis factors.}
\be
\bar \alpha_3(\bar \beta_3)-\bar \beta_3=\sum_{k=0}^n \frac{\bar \alpha_3^{(k)}}{\bar \beta_3^k}.
\label{ansatzalpha3}
\ee

\begin{figure}[t]
\centering
  \includegraphics[scale=.265]{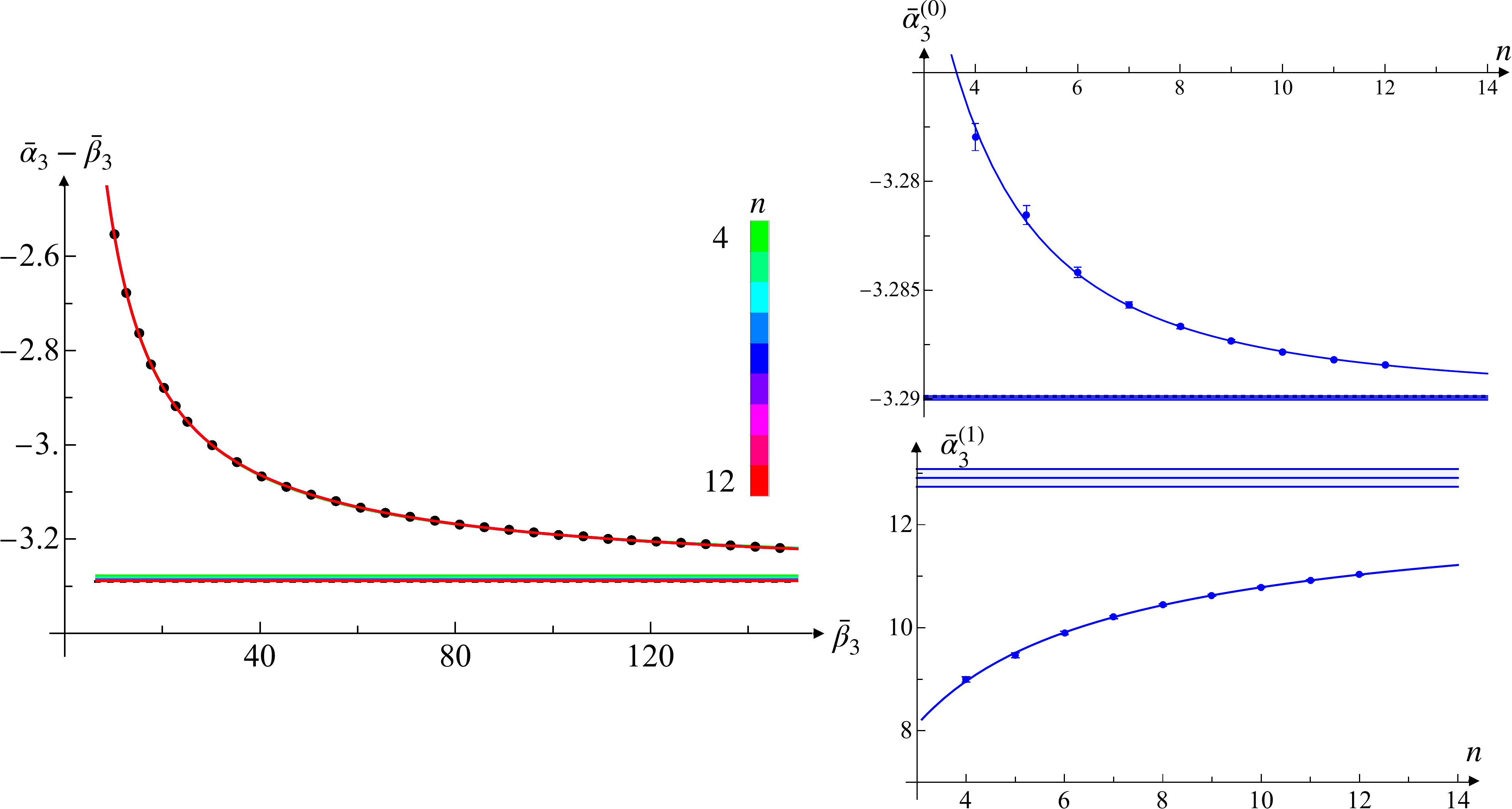}
  \caption{Left: the black dots correspond to the data for $\bar\alpha_3-\bar\beta_3$ as a function of $\bar\beta_3$; the coloured curves to the fits obtained using the ansatz in~\eqref{ansatzalpha3} for different values of the truncation $n$; the coloured horizontal lines are the extrapolated asymptotic values of  $\bar\alpha_3-\bar\beta_3$. The dashed black line is the analytic guess $-\pi^2/3$.  Right: the blue dots are the extrapolated values of $\bar\alpha_3^{(0)}$ (top), and $\bar\alpha_3^{(1)}$ (bottom) with the corresponding error bars. The blue curve is the power law Ansatz used to further extrapolate in $n$. The blue bands represent the final extrapolated values with the error bar. }
\label{fig:alpha3extrapolation}
\end{figure}

In figure \ref{fig:alpha3extrapolation} on the left, the black dots correspond to the numerical values of $\bar \alpha_3-\bar \beta_3$ as a function of $\bar \beta_3$. 
In the same figure, we plot the curves obtained by fitting the data using the ansatz in \eqref{ansatzalpha3} for different values of the truncation parameter $n$ ranging from $n=2$ in green to $n=9$ in red.
On the right, we show the first two corrections to the EFT prediction $\bar \alpha_3^{(0)}$ and $\bar \alpha_3^{(1)}$ as a function of the truncation parameter $n$.
The error bars are extracted from the fit procedure, and do not include a systematic error related to the finite $N_\text{max}$ size of our numerical Ansatz that we use for the fits. The data used are obtained for $N_\text{max}=400$. However, each observable we extract from the bootstrap has a different rate of convergence in $N_\text{max}$. For instance, the higher the order of the coefficients in the low energy expansion, the worse it converges. $\bar\alpha_3$ is the fastest to converge, and we can safely use all the points to fit.
We further extrapolate in $n$ for each of the coefficients $\bar \alpha_3^{(k)}$ using a power law Ansatz of the form $f(n)=a+b n^{-c}$, finding that \be
\bar \alpha_3(\bar \beta_3)-\bar \beta_3=-3.2900(1)+\frac{12.9(2)}{\bar \beta_3}+\mathcal{O}(\bar\beta_3^{-2})
\ee
We can conjecture that the leading correction converges to $\bar \alpha_3^{(0)}=-\pi^2/3\simeq-3.2899$. In the next section, we will explain this conjecture and it will be related to the specific UV completion of the bootstrap amplitude.
The extrapolation of the higher order corrections $\bar\alpha_3^{(k)}$ with $k>2$ requires more precision, and we do not attempt it here.

\begin{figure}[t]
\centering
  \includegraphics[scale=.265]{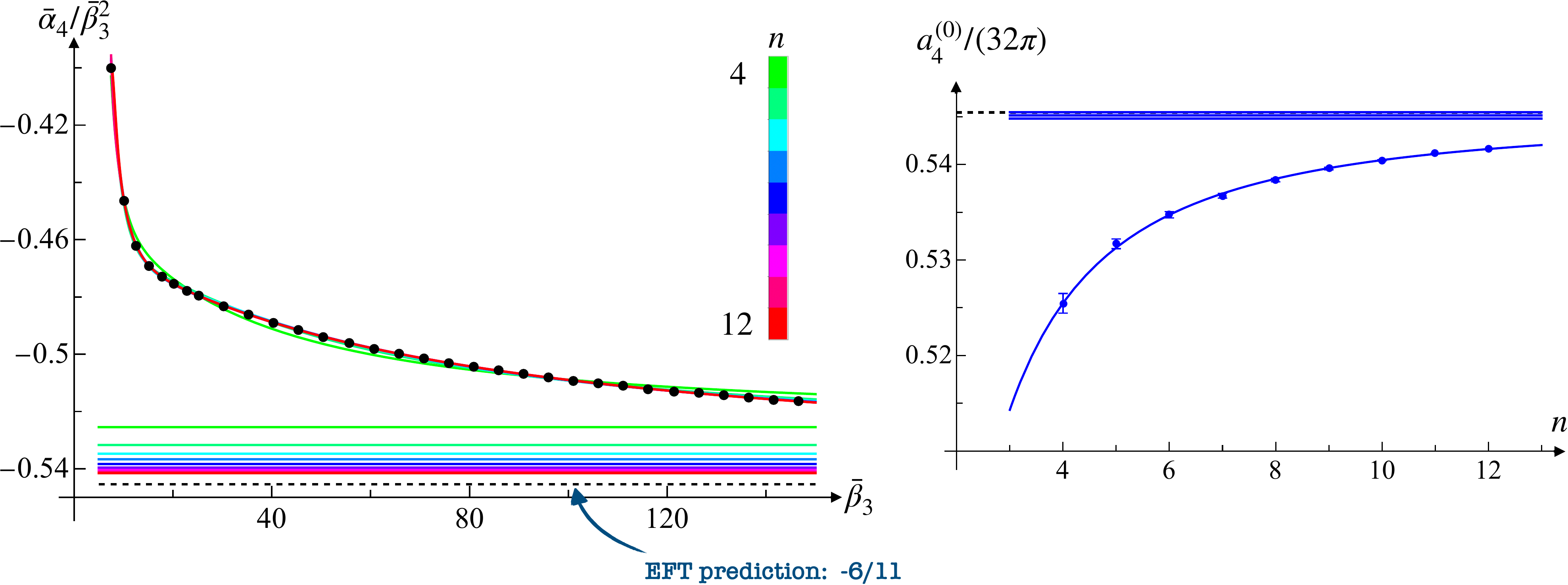}
  \caption{Left: black dots are the values of the ratio $\bar\alpha_4/\bar\beta_3^2$. We extrapolate the data using eq.~\eqref{eq:extrapolation_alpha4} for different values of $n$. Curves with different colours are obtained with different $n$. The horizontal lines represent the asymptotic value extrapolated. Right: the extrapolated value of $a_4^{(0)}$ with the error bars. The blue curve is the power law fit of the data. The blue band is the final result. The black dashed line is the EFT prediction.}
\label{fig:alpha4extrapolation}
\end{figure}

The scaling theory of the light axion also predicts that
\be
\bar\alpha_4=-\frac{a_4}{32 \pi} \bar \beta_3^2+\mathcal{O}(\bar\beta_3), \quad a_4=-\frac{1}{\alpha_2}=\frac{192\pi}{11}\simeq 54.83.
\ee
Similar to what we have done above for $\alpha_3$, we can unambigously extract the $\alpha_4$ coefficient from the low energy expansion of the numerical S-matrices obtained by minimizing $\alpha_3$ at fixed $\beta_3$.\footnote{In $D=4$, the logarithms appear in the phase shifts at order $\mathcal{O}(s^5)$. In $D>4$, they appear at the same order as $\alpha_4$, making the extractions of this coefficient from our numerical amplitudes ambiguous.}
The values obtained are shown in figure~\ref{fig:alpha4extrapolation} (the black dots). As before, we use the following parametrization to extrapolate the data for large $\bar\beta_3$:
\be
\frac{\bar\alpha_4(\bar \beta_3)}{\bar \beta_3^2}=-\frac{1}{32\pi}\sum_{k=0}^n \frac{a_4^{(k)}}{\bar \beta_3^k}.
\label{eq:extrapolation_alpha4}
\ee
In figure \ref{fig:alpha4extrapolation}, on the left, we plot the result of the various fits where each colour corresponds to a different $n$. The horizontal lines represent the extrapolated asymptotic value of $\bar\alpha_4/\bar\beta_3^2$.
In the same figure, on the right, we perform an additional extrapolation using a power law Ansatz in the truncation parameter $n$ obtaining a value $\tfrac{a_4^{(0)}}{32\pi}= 0.5451(3)$ in agreement with the analytic asymptotic prediction $\tfrac{a_4}{32 \pi}=6/11\simeq 0.5454$. 

\begin{figure}[t]
\centering
\includegraphics[scale=0.27]{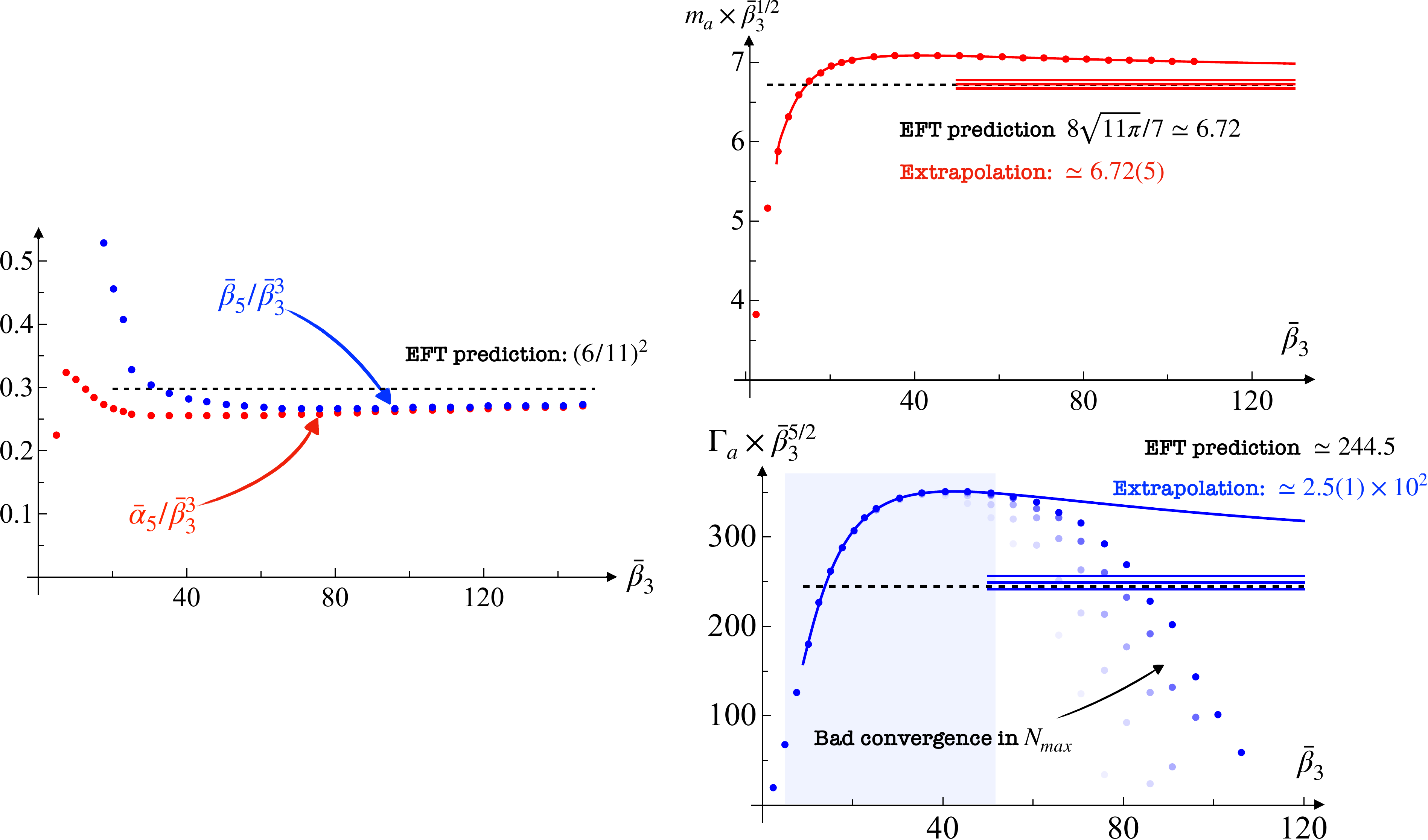}
 \caption{Left: comparison of the two ratios $\bar\alpha_5/\bar\beta_3^3$ and $\bar\beta_5/\bar\beta_3^3$ with the analytic EFT prediction. Top right: mass of the axion as a function of $\bar\beta_3$ (red dots), and the fit with an inverse power Ansatz with $n=11$ (red curve). The error band is determined by combining the results for $4\leq n\leq 11$, and further extrapolating in $n$. Bottom right: width of the axion (blue dots), and the inverse power fit with $n=7$ (blue curve). Only the points in the blue shaded region have been used for the extrapolation. The final estimate and the blue error band were obtained combining fits with $4\leq n\leq 7$.}
 \label{fig:/EFT_predictions}
\end{figure}

The predictions for the higher order coefficients are harder to check. The reason is that our numerical Ansatz does not contain logs and cannot match precisely the EFT low energy expansion at such higher orders in eq.~\eqref{4D_ps}. However, we expect that in the regime of large $\beta_3$, logs should not matter because they are parametrically suppressed. In figure~\ref{fig:/EFT_predictions} (left panel), we show respectively in blue and in red the two ratios $\bar\beta_5/\bar\beta_3^3$ and $\bar\alpha_5/\bar\beta_3^3$. 
Indeed, the coefficients $\bar\alpha_5$ and $\bar\beta_5$ are extracted by matching the low energy expansion of our ansatz with \eqref{4D_ps} setting the logs identically to zero.
We clearly see that the EFT prediction 
\be
\bar\alpha_5=\bar\beta_5=\left(\frac{6}{11}\right)^2\bar\beta_3^3+\mathcal{O}(\bar\beta_3^2)
\ee
is well verified.

\begin{figure}[t]
\centering
\includegraphics[scale=0.3]{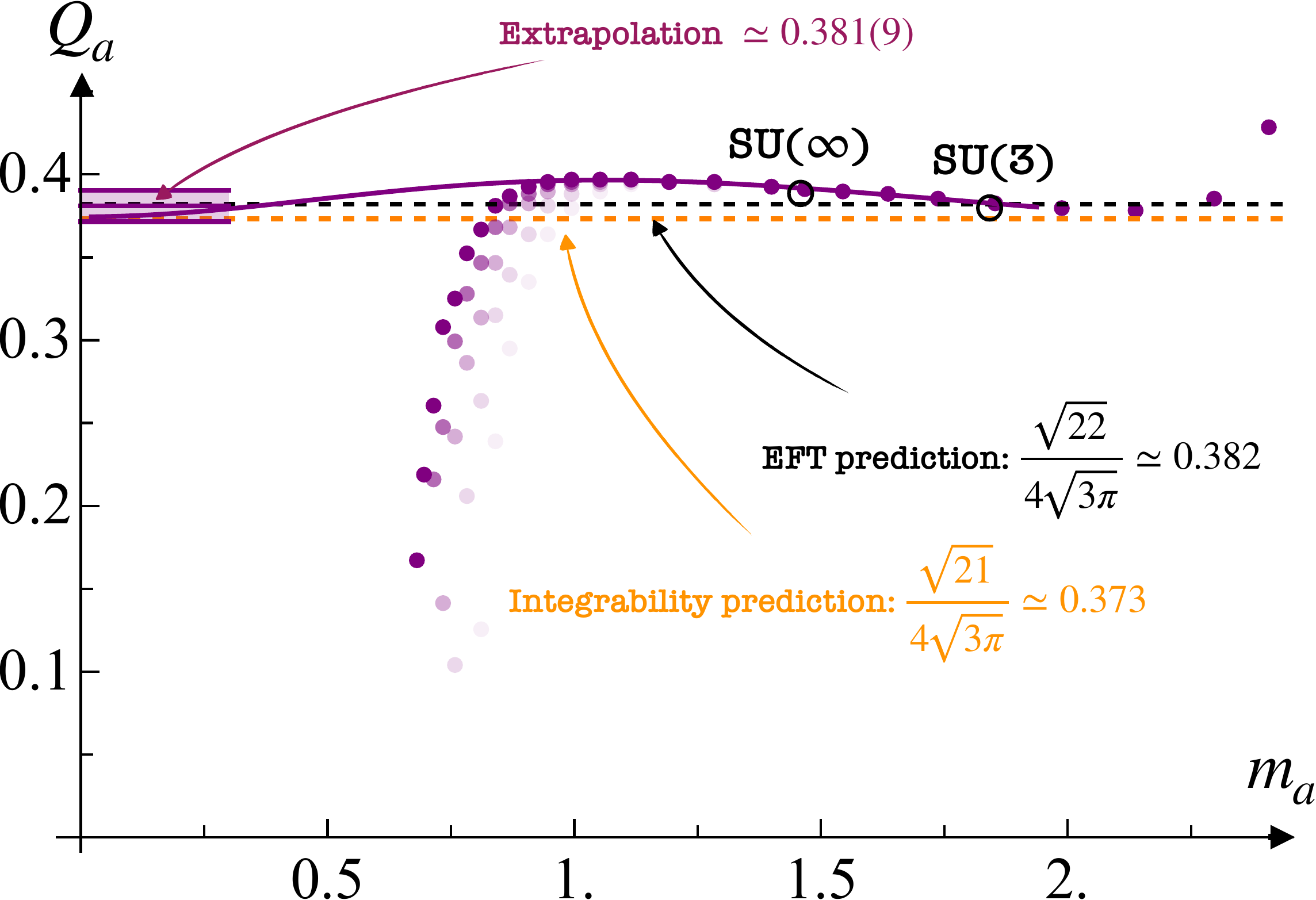}
 \caption{Nonperturbative axion charge $Q_a$ as a function of the mass $m_a$. The lightest points have $N_\text{max}=200$, the darkest ones $N_\text{max}=400$. The purple curve has been determined by fitting separately the mass and the width as in fig.~\ref{fig:/EFT_predictions} and constructing the ratio $\sqrt{8\Gamma_a m_a^{-5}}$. The band represents the extrapolated charge and its error.}
 \label{fig:MassCharge}
\end{figure}

Next, we extract the axion parameters from the zeros of the S-matrix in the antisymmetric channel
\be
S_\text{anti}(s_a)=0.
\ee
For $\bar\beta_3>0$, we always find two zeros. We call axion the zero with smaller $\re s_a$, while the heavier will be called axion*. By definition the mass and width are related to the zero position by the relation $s_a=(m_a+i\Gamma_a/2)^2$.
The results are shown in figure~\ref{fig:/EFT_predictions} on the right. In the top right panel, the red dots are the values of the mass of the axion as a function of $\bar\beta_3$. 
We perform the same extrapolation procedure already explained for $\bar\alpha_3$ and $\bar\alpha_4$ using different truncated inverse power expansions. In particular, we consider $4\leq n\leq 11$. We repeat the same analysis for the width of the axion $\Gamma_a$.
However, for $\bar\beta_3>50$ this observable still depends on $N_\text{max}$. The dots in blue, from light to dark, correspond respectively to $N_\text{max}=200$ to $N_\text{max}=400$ in steps of 50. The blue shaded region indicates the window used to extrapolate $\Gamma_a$. Given that we can use less points compared to the other cases, we extrapolate using $4\leq n\leq 7$. The final result is
\be
m_a^\text{extr}\bar\beta_3^{1/2}=6.72(5),\quad \Gamma^\text{extr}_a\bar\beta_3^{5/2}=250(10)
\ee
to be compared with the EFT predictions
\be
m_a\bar\beta_3^{1/2}=16\pi \sqrt{|\alpha_2|}=\frac{8\sqrt{11\pi}}{7}\simeq 6.72,\quad \Gamma_a\bar\beta^{5/2}=(16\pi)^5 |\alpha_2|^{7/2}\simeq 244.5.
\ee

Finally, we can use this analysis to estimate the value an error of the charge\footnote{This definition of the charge comes from matching the Lagrangian definition \eqref{ASA} to the narrow-width approximation of the axion resonance shape, below we will use it even when perturbation theory appears to break down.}
\be
Q_a^\text{extr}=\sqrt{\frac{8\Gamma_a^\text{extr}}{(m_a^\text{extr})^5}}=0.381(9).
\label{axion_charge_bootstrap}
\ee
Looking at the error propagation for the values of the mass and width extrapolated, we would expect the uncertainty to be dominated by the error on the mass 
\be
\Delta Q_a\simeq \sqrt{2.0\times 10^{-2}\Delta m^2+5.9\times 10^{-7}\Delta\Gamma^2}.
\ee 
However, it is easy to verify that with our current results the two terms in the square root are comparable.
Reducing the uncertainty on the mass is certainly possible, since the mass of the resonance depends little on $N_\text{max}\geq N_\text{critical}$.
However, as discussed in Section \ref{section_improved_numerics}, to describe the axion resonance well, and to estimate $\Gamma_a$ for large $\bar\beta_3$, we must have $N_\text{max}\gg N_\text{critical}$. The largest Ansatz we use has $N_\text{max}=400$. With this $N_\text{max}$, we find an axion until the point $\bar\beta_3=106.1$. For this value of $\bar\beta_3$, we estimate $N_\text{critical}\simeq 300$. Empirically, we have established that we need at least $N_\text{max}=3 N_\text{critical}$ to have good convergence of a resonance. In this case, we need at least  $N_\text{max}\simeq 10^3$. 
We think that designing more adaptive Ansatzes, allowing for multiple tunable scales, it might be possible to achieve a better convergence with less terms.

With the current precision, we cannot discriminate between our EFT prediction, and, for example, the value of $Q_a^c$ in the Axionic String Ansatz, which assumes that the axion is contributing to the $s^2$ term in the phase shift through loops. We remind that these numbers are given respectively by
\be
Q_a^{\text{EFT}}= \frac{\sqrt{22}}{4\sqrt{3\pi}}\simeq 0.382,\quad Q_a^c=\frac{\sqrt{21}}{4\sqrt{3\pi}}\simeq 0.373.
\label{axion_couplings}
\ee

More generally, the bootstrap allows us to study arbitrary values of $\alpha_2$. While most of our results correspond to the physical value, we checked that changing $\alpha_2$ modifies the bootstrap results according to the EFT expectations.

The following picture emerges from our numerical analysis: a common feature of our plots for the S-matrix coefficients, figures \ref{fig:alpha3extrapolation}-\ref{fig:/EFT_predictions} is that the deviation from the light-axion EFT predictions becomes significant below $\bar \beta_3 \approx 20$, or $m_a\approx1.5\ell_s^{-1}$. It is thus not valid for masses above this value. On the other hand, figure \ref{fig:MassCharge} shows that the approximate relation $Q_a^{b}=Q_a^c$ holds for significantly higher values of the mass. If it was the full story, the Figure \ref{fig:MassCharge} would still look very mysterious. In the next section, we study this non-perturbative regime and give some more justification for the coincidence.

\section{Axion dominance beyond EFT}
\label{sec:dominance}

\begin{figure}[t]
\centering
\includegraphics[scale=0.265]{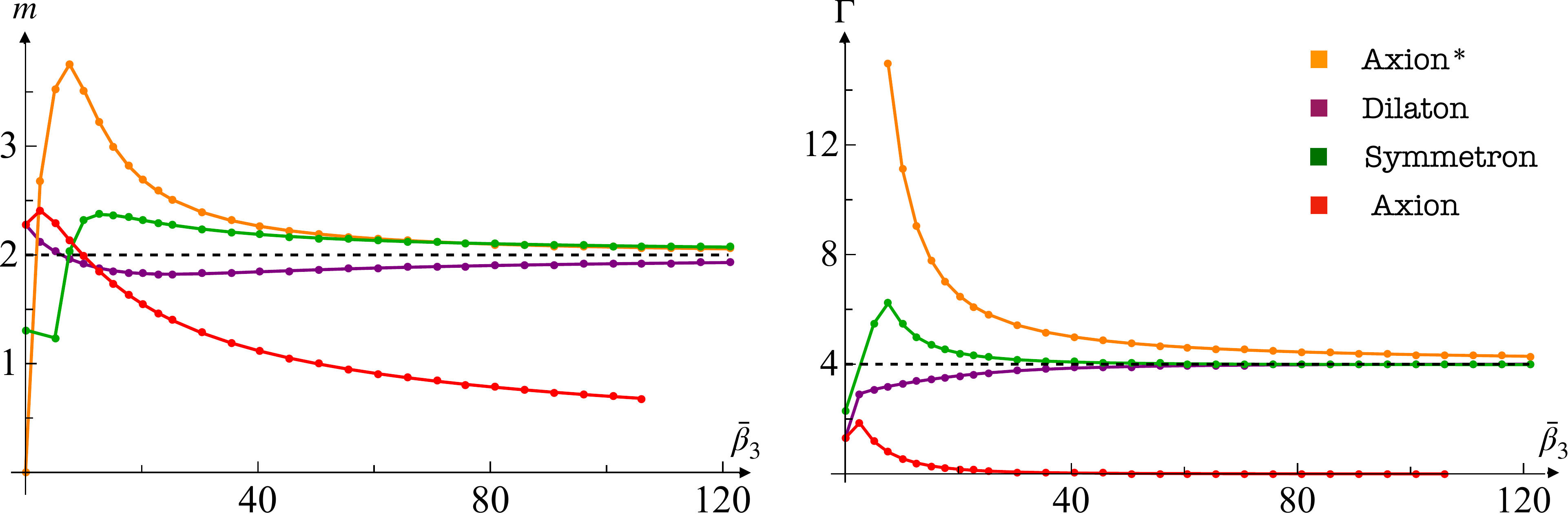}
\caption{Left: mass of all the four resonances of the S-matrix as a function of $\bar\beta_3$. Right: corresponding widths as a function of $\bar\beta_3$. Black dashed lines are the values of the mass and width of the conjectured asymptotic UV completion of the bootstrap S-matrix~\eqref{Asymptotic_S_matrix}.}
\label{fig:broad}
\end{figure}

In order to study our S-matrices beyond EFT, we first discuss
the fate of the three broad resonances present in the extremal Bootstrap amplitudes in addition to the axion. In figure \ref{fig:broad}, we show the mass and the width of these particles as a function of $\bar\beta_3$. We observe the following asymptotic behavior
\be
\lim_{\bar\beta_3\to\infty} m=2,\quad \lim_{\bar\beta_3\to\infty} \Gamma=4.
\ee
Thus, their mass does not become small in the asymptotic region and they are not present in the light-particles EFT.
Instead, we conjecture that for all channels the Bootstrap S-matrix asymptotically factorizes into a product of the form
\be
S^\text{irrep}(s)=e^{2i\delta_a^\text{irrep}} \times \frac{8i-s}{8i+s}
\label{Asymptotic_S_matrix}
\ee
Here $\delta_a^\text{anti}=\delta^\text{anti}-\frac{s}{4}$ (see \eqref{eq:asymptotic_anti}) and similarly for other channels. Thus the axion unitarizes the quadratic (PS) part of the phase shift in all channels, while the linear (NG) part is unitarized by the three broad resonances independently in each channel. One check of this conjecture is the computation of the sub-leading correction to $\bar\alpha_3$ which turns out to be, indeed, close to $-\pi^2/3$ as we observed above, see figure \ref{fig:alpha3extrapolation}.  

We also know what the S-matrix for $\beta_3=0$ is: it is an integrable solution to massless Yang-Baxter discussed in \cite{Cooper:2014noa} and \cite{EliasMiro:2019kyf}
\be
S^{\pm}_{YB}=\frac{8+is\pm32\alpha_2 s}{8-is\pm32\alpha_2 s},
\ee
where $S^+_{YB}$ is the S-matrix in the singlet and antisymmetric channel, $S^-_{YB}$ in the symmetric.
Thus the $\beta_3=0$ point in figure~\ref{fig:broad} is analytic. We see that our numerical results converge to them at smaller $\beta_3$. The S-matrix for $\beta_3=0$ has a very different nature: there are three resonances, in singlet, symmetric and anti-symmteric channels that roughly equally contribute to the unitarization of NG and PS parts of the phase shifts. The second pseudoscalar resonance simply decouples. We thus conclude that there is a transitional region between the EFT and the small $\beta_3$ regime where the axion still dominates over the other resonances, but light-axion EFT already breaks down. We will call it the \emph{axion dominance}. To study this regime we turn to sum rules and dispersion relations. We will see that indeed the axion channel gives a dominant contribution to them, explaining the value of its coupling.

The sum rules for flux tubes were developed in \cite{EliasMiro:2021nul}. Here we use the $D=4$ expressions, presenting the general $D$ formulas in the Appendix \ref{app:Sum_rules_D}. The idea is to construct crossing-symmetric combinations of amplitudes in order to reduce the integral over the real line to that of the positive half-line. In particular, the following sum rules can be constructed for the first few Wilson coefficients:
\be
0=\frac{1}{\pi}\int_0^\infty \frac{\im M_\text{sing}(z)-\im M_\text{anti}(z)}{z^3}dz
\label{zerosum}
\ee
\be
\frac{\ell_s^2}{2}=\frac{1}{\pi}\int_0^\infty \frac{\im M_\text{sym}(z)+\im M_\text{anti}(z)}{z^3}dz
\ee
\be
\alpha_2\ell_s^4=-\frac{1}{4\pi}\int_0^\infty \frac{\im M_\text{sing}(z)+\im M_\text{anti}(z)-2\im M_\text{sym}(z)}{z^4}dz
\ee
\be
\beta_3\ell_s^6=\frac{1}{4\pi}\int_0^\infty \frac{\im M_\text{anti}(z)-\im M_\text{sing}(z)}{z^5}dz
\label{beta3_sum_rule}
\ee
\be
\alpha_3\ell_s^6=\frac{1}{4\pi}\int_0^\infty \frac{\im M_\text{sing}(z)+\im M_\text{anti}(z)+2\im M_\text{sym}(z)-z^3/4}{z^5}dz+\frac{1}{384}
\ee

\begin{figure}[t]
	\centering
		\includegraphics[scale=0.265]{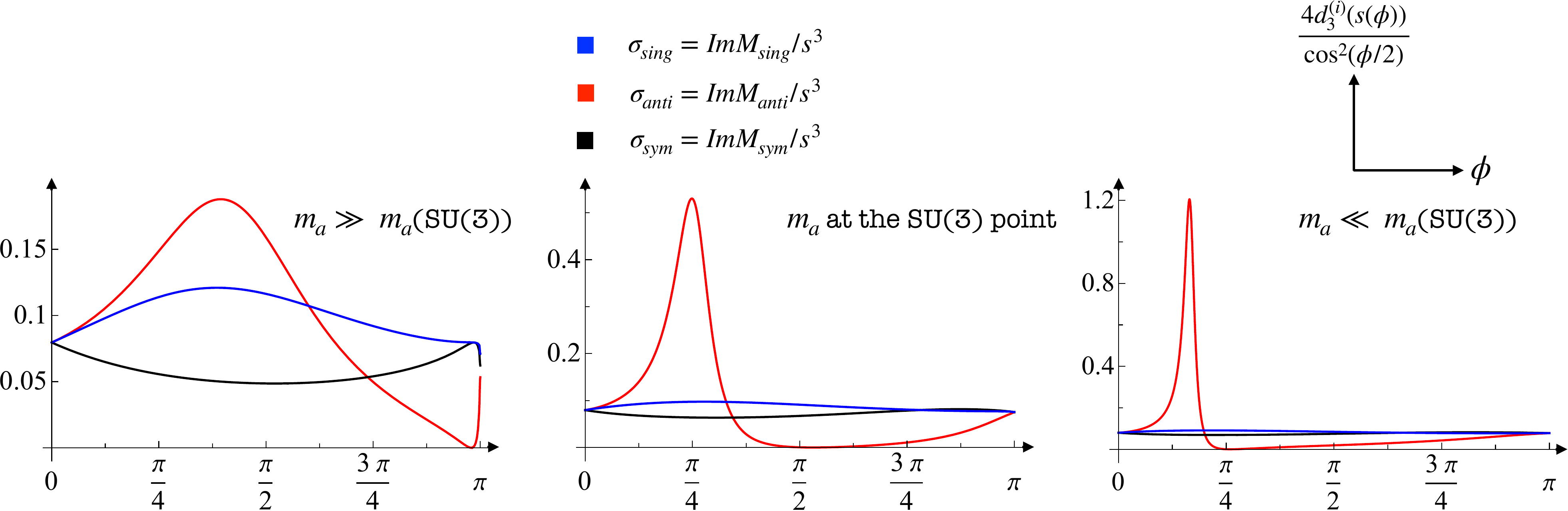}
		\caption{Integrand of the ``total cross-section'' sum rules $\sigma_\text{irrep}$ as a function of $s(\phi)=8\tan\phi/2$. The Jacobian is included in the integrand. Each irrep is highlighted by a different colour.}
		\label{fig:bound_normal}
			\label{fig:bound}
\end{figure}

First two sum rules are the ``total cross-section'' sum rules similar to the one used in \cite{Adams:2006sv}. They are related to the leading NG interactions at low energies. Higher-order sum rules are related to higher-order Wilson coeffcieints.
We can check that at tree-level they result into the same expressions as our light-axion EFT together with the UV-matching conditions. Namely, in the EFT we have $\im M_\text{sing}=\im M_\text{sym}=0$, while $\im M_\text{anti}=\tfrac{\pi}{2}g^2 m^8 \delta(s-m^2)$.
Plugging into the dispersion relations we get
\be
\beta_3\ell_s^6=\frac{1}{4\pi}\int_0^\infty \frac{\im M_\text{anti}(z)}{z^5}dz=\frac{g_a^2}{8m_a^2}
\ee
\be
\alpha_2\ell_s^4=-\frac{1}{4\pi}\int_0^\infty \frac{\im M_\text{anti}(z)}{z^4}dz=-\frac{g_a^2}{8}\,
\ee
in agreement with \eqref{UVmatch}.

Sum rules, however, can be used beyond perturbative computations. We will use them to test numerically whether the axion contribution dominates certain observables. First, note that the sum-rules related to the total cross-section are not dominated by the axion. For example, \eqref{zerosum} implies that contributions of the anti-symmetric and singlet channels must be equal. This can also be seen from figure \ref{fig:bound_normal}, where the UV part of the integrand is comparable to the axion bump. This is in agreement with the fact that the NG part of the phase shifts is unitarized by the heavy resonaces, not by the axion.

\begin{figure}[t]
	\centering
		\includegraphics[scale=0.265]{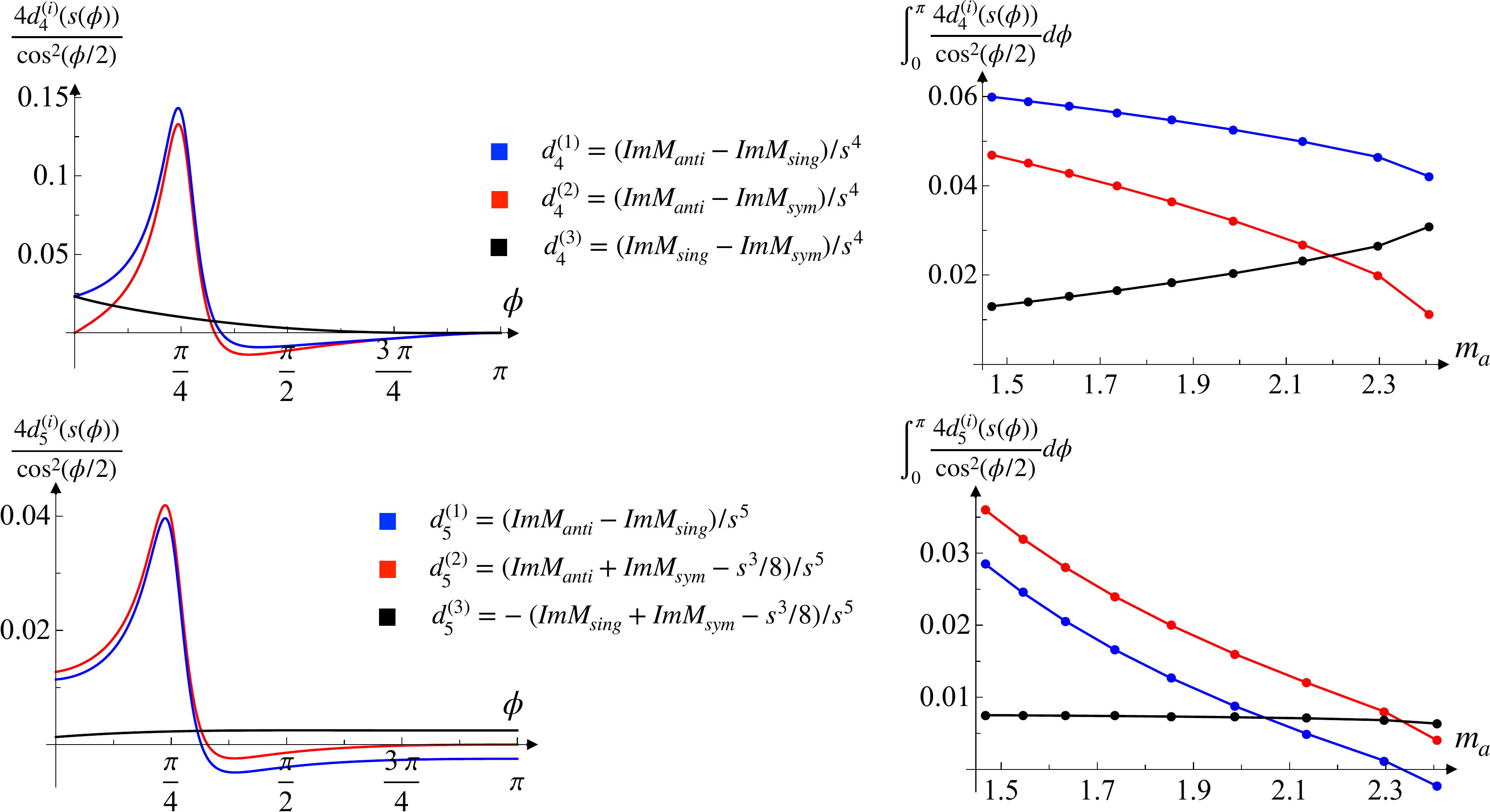}
		\caption{Left plots: integrands of IR finite dispersion relations as a function of $s(\phi)=8\tan\phi/2$ for $m_a\simeq 1.8$ (the Jacobian is included in the figure). Right plots: integrals of the IR finite dispersion relations as a function of the axion mass $m_a$.}
		\label{fig:bound_equivalent}
\end{figure}

On the other hand, higher-order sum rules are dominated by the axion. It is natural, since they are more peaked in the IR. Since integration of amplitudes in individual channels would diverge in the IR, to test this we form IR-finite linear combinations, see figure \ref{fig:bound_equivalent}. We see that for $m_a\lesssim 2\ell_s^{-1}$ the integrals that contain the anti-symmetric channel are larger than the integrals that do not. In this sense the axion dominates these dispersion relations.

\begin{figure}[t]
\centering
\includegraphics[scale=0.265]{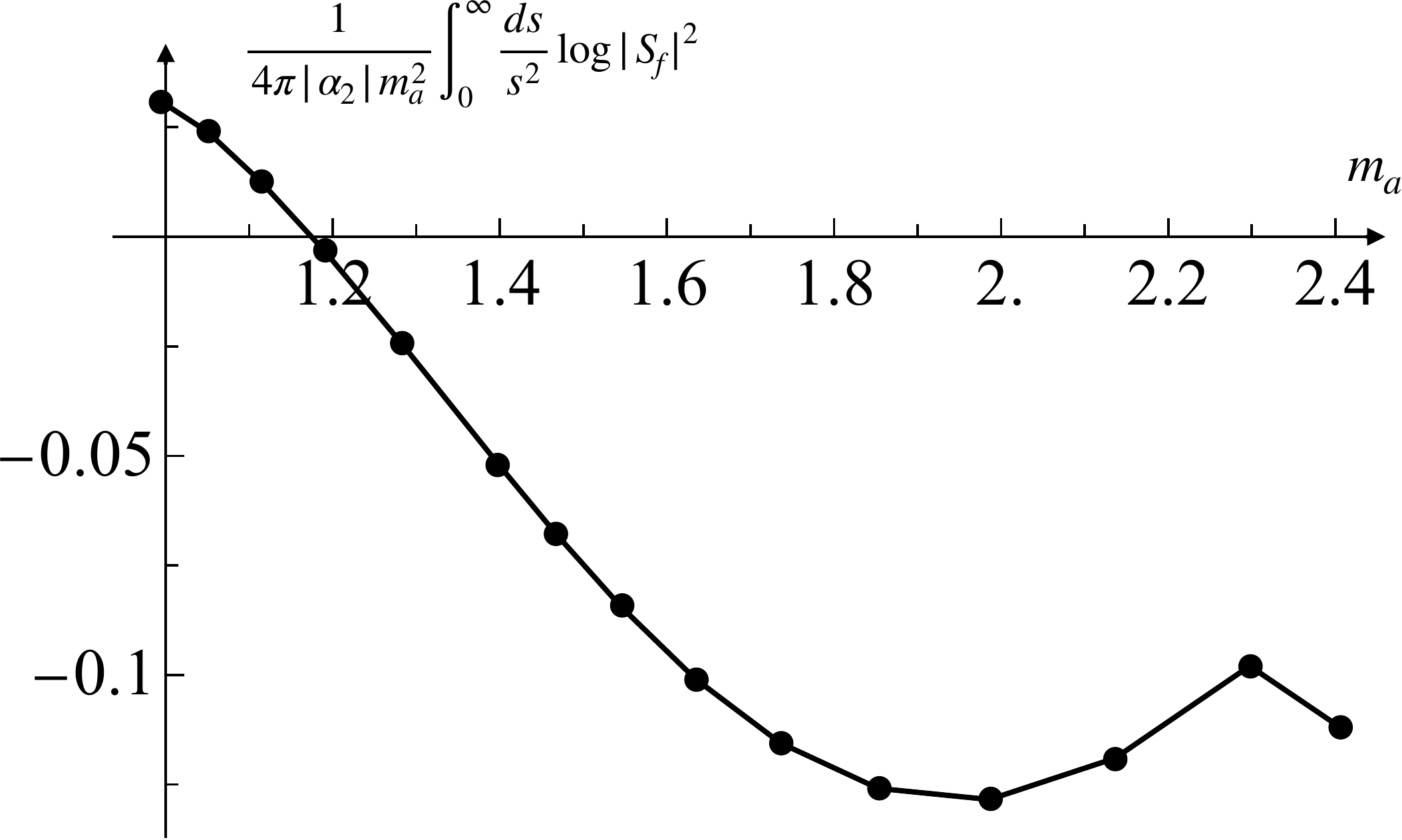}
\caption{Relative size of the integral contribution to the axion-coupling sum rule \eqref{sumrule}.}
\label{fig:axion_dominance}
\end{figure}

Finally, let us try to produce a sum rule which connects the value of $\alpha_2$ and the position of the axion zero. Here we employ the sum rule similar to the one used in \cite{Dubovsky:2018vde} for the string tension. Our logic is the following: suppose we know that there is a single narrow and relatively light resonance in the anti-symmetric channel with the real part of its mass equal to $m_a$. We would like to show that then its width should be related to $\alpha_2$. We know that the amplitude will be the largest in this channel for $s\sim m_a^2$, hence an appropriately constructed sum rule can be dominated by it. Let us consider the following object:
\be
S_f=1-\frac{ M_\text{sing}+ M_\text{anti}-2 M_\text{sym}}{2 i s^2} m_a^2\,.
\label{Sf}
\ee
It is crossing symmetric, that is $S_f(-s^*)=S_f^*(s)$. Under the assumption that $M_\text{anti}\gg M_\text{sing},M_\text{sym}$ for $s\sim m_a^2$, $S_f\sim S_{\text{anti}}$ so it will have a zero at the axion point, up to corrections of order $\Gamma/m_a\ll1$. Under the same assumptions, and also assuming that the S-matrix in the anti-symmetric channel is elastic, $|S_f|\approx 1$. We then consider the integral
\be
I_f=\oint_\gamma \frac{ ds }{s}\frac{ S_f' }{S_f}\,, 
\ee
where the contour is the same as in \cite{Dubovsky:2018vde} (line above the real axis plus the arch at infinity). Evaluating this integral by residues we get:
\be
I_f=2 \pi i \sum_{zeros} \frac{1}{s_i}\,.
\ee
On the other hand, because $\frac{ S_f' }{S_f } \approx - 4i \alpha_2 m_a^2 s $ for small $s$, evaluating the integral over the real line directly, integrating by parts and using crossing symmetry gives us
\be
I_f=-\pi i(- 4i \alpha_2 )m_a^2-\int_0^\infty\frac{ds\log  |S_f|^2}{s^2}\,.
\label{sumrule}
\ee
We know that $S_f$ has two crossing-symmetric zeros $s_\pm\approx\pm\left(m_a\pm i \frac{\Gamma}{2}\right)^2$. For small enough $m_a$ we also know that there will be no other zeros. This follows from the fact that by unitarity the amplitude in each channel is bounded, and, since in \eqref{Sf} all amplitudes are multiplied by an additional factor of $m_a^2/s$, there can be no zeros of $S_f$ when this ratio is large enough. We, however, would like to use our sum rule in the regime when other the resonances have masses comparable to the axion, so we do not have an analytic argument for the absence of other zeros. Nevertheless, we checked numerically that it is indeed the case. It would be nice to construct a sum rule for $\alpha_2$ such that the absence of zeros is more manifest, or such that their contribution is significantly suppressed.

Using the fact that $S_f$ has no other zeros and expanding to leading order in $\Gamma_a/m_a$, we get:
\be
\frac{\Gamma_a}{m_a^5}=-\alpha_2-\frac{1}{4 \pi m_a^2}\int_0^\infty\frac{ds\log  |S_f|^2}{s^2}\,.
\ee
The integral is evaluated numerically in Figure \ref{fig:axion_dominance}. We see that it gives a relative correction of order 10\% for the mass range we are interested in, since $|S_f|$ is relatively small as we expected. This results into order 5\% correction once we take the square root to extract the dimension-less charge according to \eqref{Qa}. We thus conclude that the sum rule \eqref{sumrule} explains sufficiently well why the axion coupling is close to the critical value even when the EFT breaks down, but when the axion is still light enough to dominate the dispersion relations. It is plausible that the same sum rule applies to real QCD flux tubes, and the couplings coincidence \eqref{triple} hints in this direction. To make a careful statement we need to make sure that there are no extra zeros of $S_f$.

\section{Conlusions \& outlook}

In this work, we have explored a class of flux-tube S-matrices saturating the Bootstrap bounds, and featuring a light axion resonance. 
We have shown how it is possible to analytically predict many properties of these S-matrices using the EFT of light weakly coupled particles in an asymptotic regime. To access numerically this regime and check the predictions, we have improved the numerical convergence of the primal Bootstrap by taking into account the EFT power counting.

We have begun our discussion addressing the ``triple coincidence" among Lattice, Integrability, and Bootstrap. 
It is now clear that when the axion is sufficiently light this coincidence can be explained using the EFT of light-particles.
For physical values of the axion masses, however, the EFT prescription breaks down. Nevertheless, we have found that even for 
masses close to the cutoff of the effective string description, as it is the case for $SU(N)$ YM theories, dispersion relations are dominated by the axion. We call this regime \emph{axion dominance}. 
By constructing a suitable sum-rule \eqref{sumrule}, we have proven that the axion coupling is indeed related to the PS coefficient $\alpha_2$ up to an error that turns out to be numerically small for the amplitudes that saturate the Bootstrap bounds.

In the light-particle EFT, the role of the axion is to cancel the universal PS term at intermediate energies above its mass, but below $\Lambda_{QCD}$. 
In the Axionic String, the coupling of the axion is tuned to cancel the particle production at low energies which is again proportional to the PS term.
The two mechanisms lead to similar values for the axion coupling. The small difference is due to whether or not the axion runs in the loop,
see eq.~\eqref{axion_couplings}. 
In the non-perturbative axion dominance regime, we might conjecture that the worldsheet YM axion is also responsible for an approximate cancellation of particle production in some intermediate regime, thus suppressing the inelasticity effects. If that were the case, one could hope that there exists an integrable model close to the real QCD flux tube theory. Finding such a model would give us a fundamental tool to understand the dynamics of confining strings.

Below we list a number of possible extensions of our work.

\begin{itemize}
\item 
In this work we studied only the two-particle S-matrix.
Including multi-particle processes into the Bootstrap is a notoriously hard task.\footnote{See~\cite{to_appear_multiparticles} for a first exploration in the case of $D=3$ flux tubes.}  A simpler approach is to study the sensitivity of the Bootstrap bounds to the inelasticity profile, see also~\cite{Antunes:2023irg}. It would be interesting to parametrize the inelasticity function in the $2\to 2$ scattering amplitude and determine what are the relevant features that let the bound change.
Another approach is to estimate the particle production from the lattice data with the help of dispersion relations, and inject this information into the Bootstrap. 

\item
Another source of discrepancy with the real flux tube theory comes from the tuning of higher order coefficients. We have checked that the bounds on $\alpha_3$ and $\beta_3$ do not change when we vary the value of $\alpha_4$. There is an RG intuition for this: even if we fix the lowest energy Wilson coefficients within the allowed region, there are still infinite possible consistent UV completions. However, for any fixed value of $\alpha_4$, it is possible to look for bounds in the $\{\alpha_5,\beta_5\}$ space. In general, by tuning these higher order coefficients we should expect the appearance of other light resonances in all channels. It would be interesting to study their effect on the bounds and on the axionic S-matrix. 
This would correspond to moving inside the allowed region.
Combining this analysis with the empirical absence of additional massive states on the flux-tube will help us to better establish our picture.

\item

In this work, we have started the exploration of confining strings in any $D$. In Appendix~\ref{App:genD}, we summarize the EFT predictions in any dimension. 
We have numerically checked that our predictions are explicitly verified in dimension $D=5,15$, and we find an axion (anti-symmetric form field) for $\bar \beta_3\gg1$ with the expected properties. In $D=30$, we have found a weakly coupled symmetron (a resonance in the traceless symmetric channel), but with a mass that did not become light for the range of $\bar\beta_3$ we have studied. It would be interesting to construct a similar sum-rule to the one in~\eqref{sumrule}, and check the symmetron dominance, and understanding the large $\bar\beta_3$ regime for $D>26$. 
In figure~\ref{fig:All_D_boundary}, we observe interesting things happening as we vary $D$: 
the bound on $\alpha_3$ reaches a maximum around dimension $D=9$ before going down, and at $D>26$ we have a sharp transition where the axion is replaced by the symmetron, while for $D=26$ the bounds saturate the analytic Schwarz-Pick inequalities. A more detailed study of general $D$ confining strings might elucidate the mechanism behind these phenomena. We discussed only briefly the case of negative $\bar\beta_3$. The axion in this case gets replaced by the dilaton, thus in this region of parameters bootstrap can be relevant for studying a string propagating in AdS-like spacetimes. 

\item

Dispersion relations and sum rules provide an alternative and insightful description of scattering amplitudes.
For instance, the fact that for large positive $\bar\beta_3$ the antisymmetric channel dominates over the singlet can be simply interpreted using the sum rules in eq.~\eqref{beta3_sum_rule}. In fact, all the qualitative features we observe studying the Bootstrap S-matrices can be a posteriori connected with properties of the dispersion relations. We think it will be worth using the dispersive approach more extensively to reconstruct the flux-tube S-matrix. Using the method developed in \cite{Tourkine:2021fqh}, it might even be possible to reconstruct an amplitude where particle production is dynamically generated when we break integrability with the PS term.

\item

The determination of the axion charge from lattice data is based on a fit of the phase shifts extracted using the L\"uscher method. This fit is done using a simple unitarized model for the axion phase shift, very similar to the one we have considered in this paper \eqref{eq:asymptotic_anti}. 
Obtaining reliable phase shifts from lattice is separate problem that we are not addressing in this paper. The newest lattice data will be analyzed in \cite{to_appear_lattice_axion}.
However, in this paper we have shown that the light-particle EFT description breaks down at the QCD axion mass scale. A more precise way to fit the properties of the axion would be to use the Bootstrap solutions we have found. Each extremal amplitude we find is only function of $\beta_3$. Minimizing the norm $||\delta_\text{exp}-\delta_\text{bootstrap}(\beta_3)||$ would tell us the precise value of $\beta_3$ associated to the axion using a model that obeys unitarity, crossing, and analyticity exactly, hence fixing also the non-perturbative axion coupling.

\item

Another possible extension is to consider flux tubes in supersymmetric theories, for example $\calN=1$ Yang-Mills.
The EFT for supersymmetric confining strings which includes goldstones and goldstinos has been developed in \cite{Cooper:2014noa}. It would be interesting to understand the fate of the axion in presence of spontaneously broken supersymmetry, and if any other light resonances generically appear. It is an interesting challenge both for bootstrap and for the lattice.

\item 

In this paper, we have tuned the parameter $s_0$ of the conformal map used in the Bootstrap to optimally describe the axion for large $\bar\beta_3$. However, we have shown that there exists an intrinsic lower bound on the resolution we need to achieve, which is given by the ratio $\Gamma/m$. 
In the future, it might be worth to explore different maps, in particular, conformal maps that contain more than one scale parameter. 
For instance, a generalization of the \emph{wavelet} idea introduced in \cite{EliasMiro:2022xaa}. We believe that studying a different class of conformal maps could lead to a better performing Ansatz, and eventually reduce the uncertainty on the extrapolation of the asymptotic axion charge~\eqref{axion_charge_bootstrap}.

\end{itemize}

\section*{Acknowledgements}
We thank Sergei Dubovsky, Guzman Hernandez-Chifflet for useful discussions. 
AG is supported by the Israel Science Foundation (grant number 1197/20).
VG is supported by grant 994302 as part of the Simons Collaboration on
Confinement and the QCD String. 
Research at the Perimeter Institute is supported in part by the Government of Canada through
NSERC and by the Province of Ontario through MRI.
ALG is supported by the European Union -
NextGenerationEU, under the programme Seal of Excellence@UNIPD, project acronym CluEs. 
We are also thankful to the organizers and the participants of the KITP program ``Confinement, Flux Tubes, and Large N'' where this work was initiated. This research was supported in part by grant NSF PHY-1748958 to the Kavli Institute for Theoretical Physics (KITP).

\appendix

\section{Flux Tubes in $D$-dimensions}
\label{App:genD}
In this appendix, we study the generic Flux Tube EFT in $D$ dimensions.
As we discussed in Section~\ref{sec:EFT4D}, in $D=4$ when the sign of $\beta_3$ is positive the EFT contains a weakly coupled light axion, when $\beta_3$ is negative we find a dilaton.
To study the general case, it is convenient to couple particles in all different representations to the world-sheet branons. To be consistent with non-linear Lorentz symmetry we must consider a four derivative interactions
\be
\frac{1}{2}\partial_\alpha \partial_\gamma X^i \partial_\beta\partial^\gamma X^j(\delta^{\alpha\beta}(g_0 \mathbb{P}_0+g_2 \mathbb{P}_2)_{ij}^{kl}+\varepsilon^{\alpha\beta} g_1 (\mathbb{P}_1)_{ij}^{kl}) \Phi_{kl}\equiv\frac{1}{2}\partial_\alpha \partial_\gamma X^i \partial_\beta\partial^\gamma X^j (\mathbb{G}^{\alpha\beta})_{ij}^{kl} \Phi_{kl},
\ee
where the irreps projectors are explicitly given by
\begin{eqnarray}
&&(\mathbb{P}_0)_{ij}^{kl}=\frac{1}{D-2}\delta_{ij}\delta^{kl},\\
&&(\mathbb{P}_1)_{ij}^{kl}=\frac{1}{2}(\delta_{i}^{k}\delta_j^l-\delta_{i}^{l}\delta_j^k),\\
&&(\mathbb{P}_2)_{ij}^{kl}=\frac{1}{2}(\delta_{i}^{k}\delta_j^l+\delta_{i}^{l}\delta_j^k)-\frac{1}{D-2}\delta_{ij}\delta^{kl}.
\end{eqnarray}
We can use the Feynman rule for the interaction vertex $-i p^i_\alpha p^j_\beta (\mathbb{G}^{\alpha\beta})_{ij}^{kl}$, and the identity $\varepsilon^{\alpha\beta}p_\alpha q_\beta=-p\cdot q$ to readily compute the amplitude with generic polarizations of the external legs
\be
i \mathcal{M}_{ij}^{kl}=i M_1\delta_{ij}\delta^{kl}+i M_2\delta_{i}^{k}\delta_j^l+i M_3 \delta_{i}^{l}\delta_j^k =-i\sum_{I=\{0,1,2\}}g^2_I  (\mathbb{P}_I)_{ij}^{kl}\frac{s^4}{s-m^2_I}+i\sum_{I=\{0,1,2\}}g^2_I (-1)^I (\mathbb{P}_I)_{il}^{jk}\frac{s^4}{s+m^2_I}
\ee
From this equation, we can extract the components $M_i$
\begin{eqnarray}
&&M_1=\frac{s^4}{2}\left( \frac{2g_0^2}{(D-2)(m_0^2-s)} -\frac{g_1^2}{s+m_1^2}+g_2^2\left( \frac{1}{m_2^2+s}-\frac{2}{(D-2)(s-m_2^2)} \right)\right)\\
&&M_2=\frac{s^4}{(m_1^4-s^2)(m_2^4-s^2)}(g_2^2 m_2^2(s^2-m_1^4)+g_1^2 m_1^2(s^2-m_2^4))\\
&&M_3=\frac{s^4}{2}\left( \frac{2g_0^2}{(D-2)(s+m_0^2)} +\frac{g_1^2}{s-m_1^2}+g_2^2\left( \frac{1}{s+m_2^2}+\frac{2}{(D-2)(s-m_2^2)} \right)\right)
\end{eqnarray}
and combine them into irreps
\begin{eqnarray}
&&M_\text{sing}=(D-2)M_1+M_2+M_3\\
&&M_\text{anti}=M_2-M_3\\
&&M_\text{sym}=M_2+M_3
\end{eqnarray}
Below, we adapt these formulas to specific examples.

\subsection{Axion branch: $D<26$, $\beta_3>0$}
\label{App:anit-sym}
Our numerical results suggest that for any $D<26$ and for $\beta_3>0$, the asymptotic Bootstrap bounds are dominated by an axionic (antisymmetric form field) EFT. So, if we set $g_0=g_2=0$, and call $g_1=g$, $m_1=m$, we get
\begin{eqnarray}
&&M_\text{sing}=-(D-3)\frac{g^2}{m^4}\frac{s^4}{s+m^2}\\
&&M_\text{anti}=\frac{g^2}{m^4}\frac{s^4(s+3m^2)}{m^4-s^2}\\
&&M_\text{sym}=\frac{g^2}{m^4}\frac{s^4}{s+m^2}
\end{eqnarray}
The tree-level matching implies
\begin{eqnarray}
\alpha_3=\beta_3,\quad \alpha_4=-a_4 \beta_3^2,\quad m=\frac{1}{\ell_s \sqrt{\beta_3}\sqrt{a_4}}, \quad g=\frac{\sqrt{2}}{\beta_3 a_4^{3/2}}.
\end{eqnarray}
The predictions for the higher order coefficients are
\be
\alpha_5=\beta_5=a_4^2\beta_3^3, \quad \alpha_6=-a_4^3\beta_3^4,
\ee
the mass and charge of the axion are given respectively by
\be
m=\frac{1}{\ell_s \sqrt{\beta_3 |a_4|}}+\frac{i}{2\ell_s \beta_3^{5/2}|a_4|^{7/2}}, \quad Q_a/\ell_s^2=\frac{2\sqrt{2}}{\sqrt{|a_4|}}.
\ee
Finally, UV causality of the phase shift implies
\be
a_4=\frac{1}{\alpha_2(D)}\implies Q_a(D)/\ell^2=\frac{\sqrt{26-D}}{4\sqrt{3 \pi}}.
\ee

\subsection{Dilaton branch: $D<26$, $\beta_3<0$}
\label{dilaton_appendix}

When $\beta_3<0$, we can conjecture dynamics being dominated by a single dilaton.
Under this hypothesis we have
\begin{eqnarray}
&&M_\text{sing}=\frac{g^2}{m^4}s^4\left(\frac{1}{m^2-s}+\frac{1}{(d-2)(m^2+s)}\right)\\
&&M_\text{anti}=-\frac{g^2}{m^4}\frac{s^4}{(d-2)(s+m^2)}\\
&&M_\text{sym}=\frac{g^2}{m^4}\frac{s^4}{(d-2)(s+m^2)}
\end{eqnarray}

The tree-level matching implies
\begin{eqnarray}
&&\alpha_3=-\beta_3, \quad \alpha_4= -a_4 \beta_3^2,\quad m=\frac{1}{\ell_s \sqrt{|\beta_3|}\sqrt{a_4}},\quad g=\frac{\sqrt{2(D-2)}}{|\beta_3|a_4^{3/2}}.
\end{eqnarray}
Predictions for the higher order coefficients are
\be
\alpha_5=-\beta_5=a_4^2|\beta_3|^3, \quad a_6=a_4^3\beta_3^4,
\ee
and for the mass and charge of the dilaton
\be
m_d=\frac{1}{\ell_s \sqrt{|\beta_3| a_4}}+(D-2)\frac{i}{4\ell_s |\beta_3|^{5/2}a_4^{7/2}}, \quad Q_d/\ell_s^2=\frac{2\sqrt{D-2}}{\sqrt{|a_4|}}.
\ee
Imposing UV causality, we get
\be
|a_4|=-\frac{1}{\alpha_2(D)},\quad Q_d(D)/\ell_s^2=\frac{\sqrt{(D-2)(26-D)}}{4\sqrt{6 \pi}}.
\ee

\subsection{Sum-rules in any $D$}
\label{app:Sum_rules_D}

In general $D$, we can use the crossing symmetric components introduced in~\cite{EliasMiro:2021nul} to write down sum rules for the imaginary part of the flux-tube scattering amplitudes
\be
0=\int_0^\infty \frac{2\im M_\text{sing}(z)-(D-2)\im M_\text{anti}(z)+(D-4)\im M_\text{sym}(z)}{z^3}dz,
\ee
\be
\frac{\ell_s^2}{2}=\frac{1}{\pi}\int_0^\infty \frac{\im M_\text{sym}(z)+\im M_\text{anti}(z)}{z^3}dz,
\ee
\be
\alpha_2\ell_s^4=-\frac{1}{4(D-2)\pi}\int_0^\infty \frac{2\im M_\text{sing}(z)+(D-2)\im M_\text{anti}(z)-D\im M_\text{sym}(z)}{z^4}dz,
\ee
\be
\beta_3\ell_s^6=-\frac{1}{4(D-2)\pi}\int_0^\infty \frac{2\im M_\text{sing}(z)-(D-2)\im M_\text{anti}(z)+(D-4)\im M_\text{sym}(z)}{z^5}dz,
\ee
\be
\alpha_3\ell_s^6=\frac{1}{4(D{-}2)\pi}\int_0^\infty \frac{2\im M_\text{sing}(z)+(D{-}2)\im M_\text{anti}(z)+(3D{-}8)\im M_\text{sym}(z)-(D-2)\tfrac{z^3}{4}}{z^5}+\frac{1}{384}.
\ee
\newpage
\section{Table with the $D=4$ data}
\label{appendix_table}
\begin{table}[h!]
\centering
\begin{tabular}{|  c || c | | c || c ||  c || c || c || c || c |} 
 \hline
 $\bar \beta_3$& $\bar \alpha_3$ & $\bar \alpha_4$ & $\bar \alpha_5$ & $ \bar \beta_5$ & $m_a$ & $\Gamma_a$ & & \\ [0.5ex] 
 \hline 
 \hline
 $ 2.527$ & 1.085 &  6.624 & -24.52 & 520.9 & 2.408 & 0.9305 & & \\
  \hline
 $ 5.053$ & 3.007  & -5.580 & 28.98  & 706.6 & 2.298 & 0.5931 & & \\
 \hline
 $ 7.580$ & 5.217 & -45.59 & 140.8 & 955.8 & 2.136 & 0.3978 & &  \\
 \hline
 $ 10.11$ & 7.554 & -22.98 & 321.9 & 1283 & 1.986 & 0.2780 & &  \\
 \hline
 $ 12.63$ & 9.956 & -73.76 & 598.1 & 1705 & 1.853 & 0.1994 & & \\
 \hline
 $ 15.16$ & 12.40 & -107.8 & 991.2 & 2248 & 1.736 & 0.1462 & & \\
 \hline
 $ 17.69$ & 14.86 & -147.9 & 1508 & 2925 &  1.634 & 0.1095 & &\\
 \hline
 $ 20.21$ & 17.33 & -194.3 & 2197 & 3757 & 1.545 & $8.363\times 10^{-2}$ & &\\
 \hline
 $ 22.74$ & 19.82 & -247.0 & 3075 & 4787 & 1.467 & $6.512\times 10^{-2}$ & &\\
 \hline
 25.27 & 22.31 & -306.2 & 4164 & 5300 &  1.398 & $5.160\times 10^{-2}$ & &\\
 \hline
 30.32 & 27.32 & -444.1 & 7127 & 8490 & 1.283 & $3.390\times 10^{-2}$ & &\\
 \hline
 35.37 & 32.33 & -608.3 & $1.129\times 10^{4}$ & $1.288\times 10^4$ & 1.191 & $2.344\times 10^{-2}$ & &\\
  \hline
 40.43 & 37.36  & -799.2 & $1.684\times 10^4$ & $1.865\times 10^4$ &1.114  & $1.688\times 10^{-2}$ & &\\
   \hline
 45.48 & 42.39 & -1017 & $2.398 \times 10^{4}$ & $2.601\times 10^4$ & 1.050 & $1.258\times 10^{-2}$ & &\\
  \hline
 50.53 & 47.43 & -1261 & $3.292\times 10^4$ & $3.519\times 10^4$ & 0.9959 & ${\color{blue}9.619\times 10^{-3}}$ & &\\
  \hline
 55.59 & 52.46 &-1533 & $4.391\times 10^4$ & $4.641\times 10^4$ & 0.9487 & ${\color{blue}7.493\times 10^{-3}}$ & &\\
  \hline
 60.64 & 57.51 & -1831 & $5.708\times 10^4$ & $5.981\times 10^4$ & 0.9075 & ${\color{blue}5.912\times 10^{-3}}$ & &\\
  \hline
 65.69 & 62.55 & -2157 & $7.275\times 10^4$ & $7.570\times 10^4$ & 0.8710 & ${\color{blue}4.670\times 10^{-3}}$ & &\\
  \hline
 70.75 & 67.59 & -2510& $9.118\times 10^4$ & $9.436\times 10^4$ & 0.8384 & ${\color{blue}3.752\times 10^{-3}}$ & &\\
  \hline
 75.80 & 72.64 & -2890& $1.125\times 10^5$ & $1.159\times 10^5$ & 0.8092 &  ${\color{blue}2.926\times 10^{-3}}$& &\\
  \hline
 80.85 & 77.68 & -3298& $1.373\times 10^5$ & $1.409\times 10^5$ & 0.7827 &  ${\color{blue}2.284 \times 10^{-3}}$ & &\\
  \hline
 85.91 & 82.73 & -3732& $1.653\times 10^5$ & $1.691\times 10^5$ & 0.7586 & ${\color{blue}1.666 \times 10^{-3}}$ & &\\
  \hline
 90.96 & 87.78 & -4195& $1.967\times 10^5$ & $2.008\times 10^5$ & 0.7366 & ${\color{blue}1.281 \times 10^{-3}}$ & &\\
  \hline
 96.01 & 92.83 &  -4684& $2.321\times 10^5$ & $2.364\times 10^5$ & 0.7163 & ${\color{blue}7.959 \times 10^{-4}}$  & &\\
  \hline
 101.1 & 97.87 &  -5201& $2.719\times 10^5$ & $2.764\times 10^5$ & 0.6975 &  ${\color{blue}4.947 \times 10^{-4}}$ & &\\
  \hline
 106.1 & 102.9 & -5745& $3.156\times 10^5$ & $3.203\times 10^5$ & 0.6802 &  ${\color{blue}2.557 \times 10^{-4}}$  & &\\
  \hline
 111.2 &108.0  & -6317& $3.640\times 10^5$ & $3.690\times 10^5$ & & & &\\
  \hline
 116.2 & 113.0 & -6917 & $4.170\times 10^5$ & $4.222\times 10^5$ & & & &\\
  \hline
 121.3 & 118.1 & -7544& $4.759\times 10^5$ & $4.813\times 10^5$ & & & &\\
  \hline
 126.3 & 123.1 & -8198 & $5.394\times 10^5$ & $5.450\times 10^5$ & & & &\\
  \hline
 131.4 & 128.2 & -8880 & $6.080\times 10^5$ & $6.139\times 10^5$ & & & &\\
  \hline
 136.4 & 133.2 & -9589 & $6.819\times 10^5$ & $6.880\times 10^5$ & & & &\\
  \hline
 141.5 & 138.3 & $-1.033\times 10^4$& $7.620\times 10^5$ & $7.683\times 10^5$ & & & &\\
\hline
146.5 & 143.3 & $-1.109\times 10^4$& $8.490\times 10^5$ & $8.556\times 10^5$ & & & &\\

 \hline
\end{tabular}
\caption{The quantities listed in the table are extracted from the S-matrix obtained by solving the optimization problem of minimizing $\bar\alpha_3$ at fixed $\bar\beta_3$ for $N_\text{max}=400$. The black data have been used in the fits and extrapolations in section~\ref{sec:detailed_results}. We denote in blue the values of $\Gamma_a$ that still haven't converged for the value of $N_\text{max}$ used. }
\label{}
\end{table}
\newpage

\small

\bibliographystyle{utphys}
\bibliography{biblio}

\end{document}